\documentclass[prd,preprint,eqsecnum,tightenlines,nofootinbib,amsmath,amssymb,
floatfix]{revtex4}
\usepackage{graphicx}
\usepackage{subfigure}
\usepackage{color}
\usepackage{cancel}

\begin{document}

\title{Exploring the Phase Diagram with Taylor Series: Epic Voyage or Just Another Bad Trip}
\author{Mark Abraao York, Guy D. Moore}
\affiliation{McGill University Department of Physics\\
3600 Rue University\\
Montreal, QC\\
H3A 2T8}
\date{June 2011}
\begin{abstract}
It has been suggested in the literature that it may be possible to locate the
QCD critical end
point using the Taylor series of thermodynamic variables about the $\mu=0$
axis.
Since the phase transition at the critical end point is believed to be in the 3D
Ising universality class,
it would seem natural to test this method with the Ising Model, for which the
answer is already known. The finding is that it is in fact possible to
pinpoint the location of the Ising critical point using Taylor coefficients.
\end{abstract}
\maketitle

\section{Introduction}
The existence and location of a critical end point (CEP) on the QCD phase
diagram is an issue of active debate in the theoretical
\cite{Kogut,Stephanov,Halasz,Brown,Hatta,Pisarski,Gavin,Forcrand} and experimental
\cite{Abreau,Appelshauser,Agakichiev1,Agakichiev2} physics community.
Perturbation theory tells us that at sufficiently high temperatures QCD matter
exists in
a deconfined quark gluon plasma (QGP) phase \cite{Letessier,Gross,Czakon}. Since
we are well familiar with
the low temperature hadronic matter phase \cite{Wilson}, at some point on the
$\mu-T$
plane a phase transition or at least a crossover must occur.

\begin{figure}
\centering
\subfigure[\label{fig:QCDPhaseDiagram}$~$QCD]
{\includegraphics[scale=0.7]{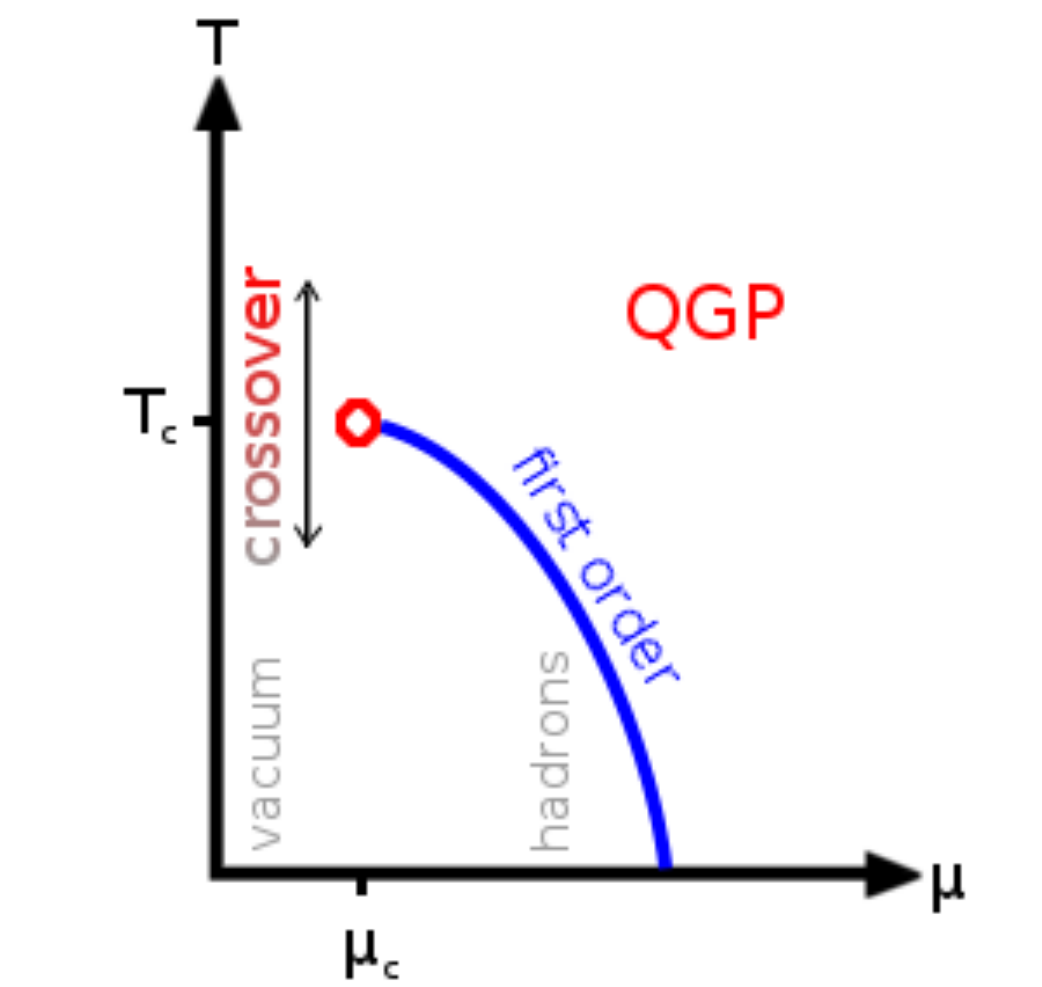}}
\subfigure[\label{fig:IsingPhaseDiagram}$~$Ising Model]
{\includegraphics[scale=0.7]{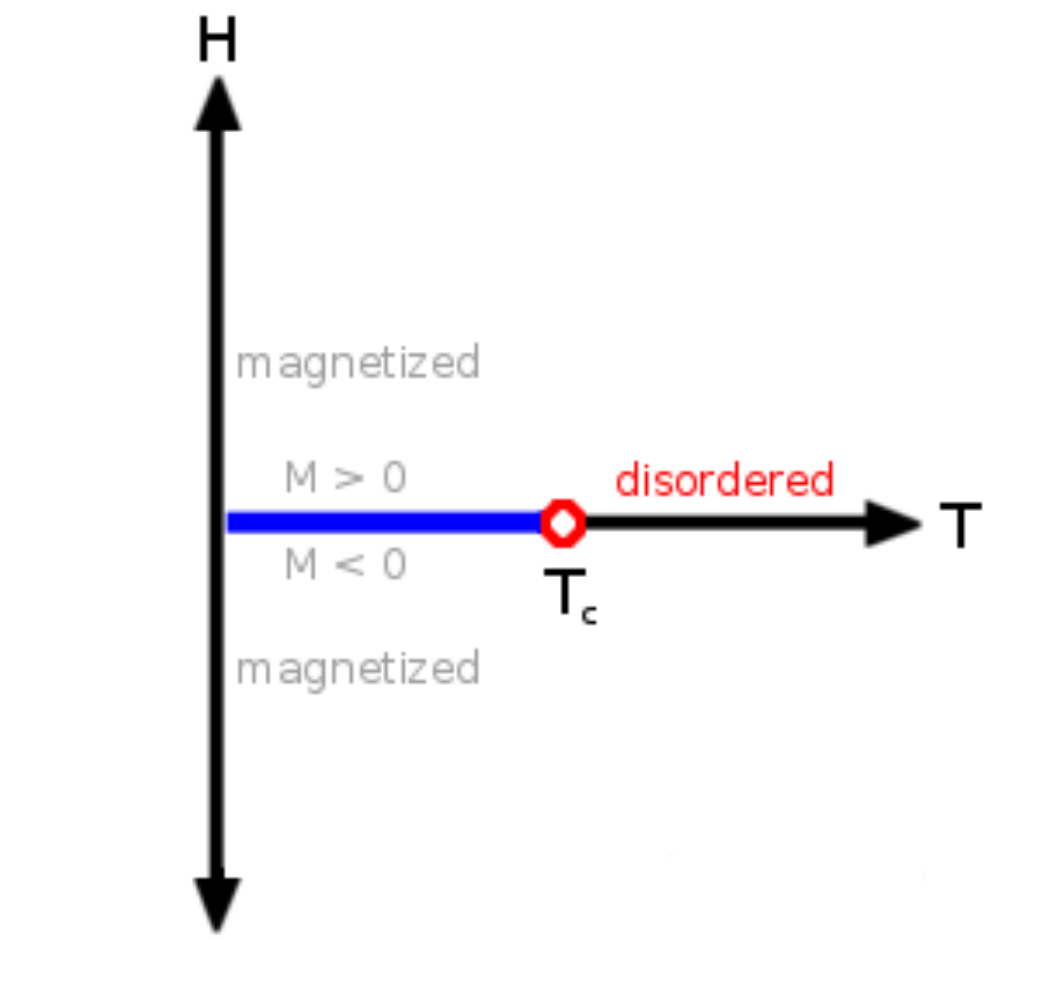}}
\caption{
On the left, an abridged version of the QCD phase diagram for physically
relevant values of $m_s$ and $m_{u,d}$, showing the existence of a CEP.
On the right, Ising Model phase diagram when $\text{D}\geq 2$.}
\end{figure}

The story is as follows, which begins with Fig. \ref{fig:QCDPhaseDiagram}: a
plausible sketch of the QCD phase
diagram for physical values of
strange, up and down quark masses.
The generic features are a first order phase boundary that terminates at a CEP
$(\mu_c,T_c)$, and a crossover between the QGP and hadronic phases at $\mu=0$. A
brief
remark about the notation: $T_c$ refers to the temperature at the CEP, while
$T_{c/o}$ denotes the QCD crossover temperature at $\mu=0$.

A first order phase boundary that terminates at a CEP is known to exist for
exactly solvable random matrix models that have the same
symmetries as QCD \cite{Klein}. Furthermore, the data that we
have
obtained from lattice QCD \cite{Lombardo,Learmann} suggests that we should
expect to see a crossover to the QGP along the $\mu=0$ axis at $T_{c/o}\sim 150
\text{ MeV}$ \cite{Aoki}. Unfortunately, it is very difficult to test these predictions experimentally, hence determining the exact layout of the QCD phase diagram remains an open problem.

For instance, in place of Fig. \ref{fig:QCDPhaseDiagram}, we may have a phase diagram with a first order curve that continues all the way to the $\mu=0$ axis. Or, alternatively, a first order phase boundary and CEP may simply not even exist. It is known theoretically \cite{Philipsen} that the transition (whether it be first order,
second order or simply a crossover) to the QGP at $\mu=0$ is heavily dependent
upon, amongst other things, the strange quark mass $m_s$. 

Assuming that the QCD phase diagram is that as depicted in Fig.
\ref{fig:QCDPhaseDiagram}, then it would be of great interest to make a
theoretical prediction about the location of the CEP. In principle, this can be
achieved via Lattice QCD by a Monte Carlo sampling of the QCD partition function
\begin{equation}                           
\label{eq:QCD_Det_Part_Func}
 \mathcal{Z}_{\text{QCD}}(T,\{\mu_f\}) = \int \mathcal{D} U e^{-S_G} \prod_f
\text{Det} M_f(\mu_f) ,
\end{equation}
however, it is well known that at non-zero chemical potential one encounters
the notorious fermion sign problem.

In spite of this limitation, the location of the CEP may still be potentially determined by lattice methods \cite{Gavai1,Ejiri,Gavai}
if one performs a Taylor expansion of thermodynamic variables about $\mu =
0$ (or more generally, $\text{Re}(\mu)=0$). Since all
of the Taylor coefficients are evaluated at zero or purely imaginary $\mu$,
there is no fermion sign problem.

Since thermodynamic variables (pressure and its derivatives for instance) are
non-analytic at the CEP, the radius of convergence of such a Taylor series
specifies the exact distance from the series' origin to the nearest non-analytic
point. If this point happens to be on the real $\mu$ axis, one has effectively
located the CEP.

The main drawback is that in practice, one is only able to compute Taylor
coefficients up to a specific finite order. Locating
the CEP, nevertheless, requires knowledge of the full Taylor series. One
would hope that it is sufficiently reliable to simply extrapolate from the few
knowable lowest order coefficients.

The focus of this work is to test the effectiveness of the Taylor series
method.
If it has any hope of predicting the location of the CEP in QCD, it
should at least be able to do so for a much simpler toy model\footnote{See also \cite{KSWW}.} whose second order
phase transition is in the same universality class. We are, of course, talking
about the 3D Ising Model, whose phase diagram is shown in Fig.
\ref{fig:IsingPhaseDiagram}.

The 3D Ising Model is not exactly solvable, but numerical simulations are able
to yield high precision estimates of its critical exponents and critical
temperature $T_c$ (unlike for QCD). I.e., the location of the Ising Model
critical point is
\textit{known}. 

\begin{figure}
\centering
\subfigure[$~$Ising]
{\includegraphics[scale=0.7]{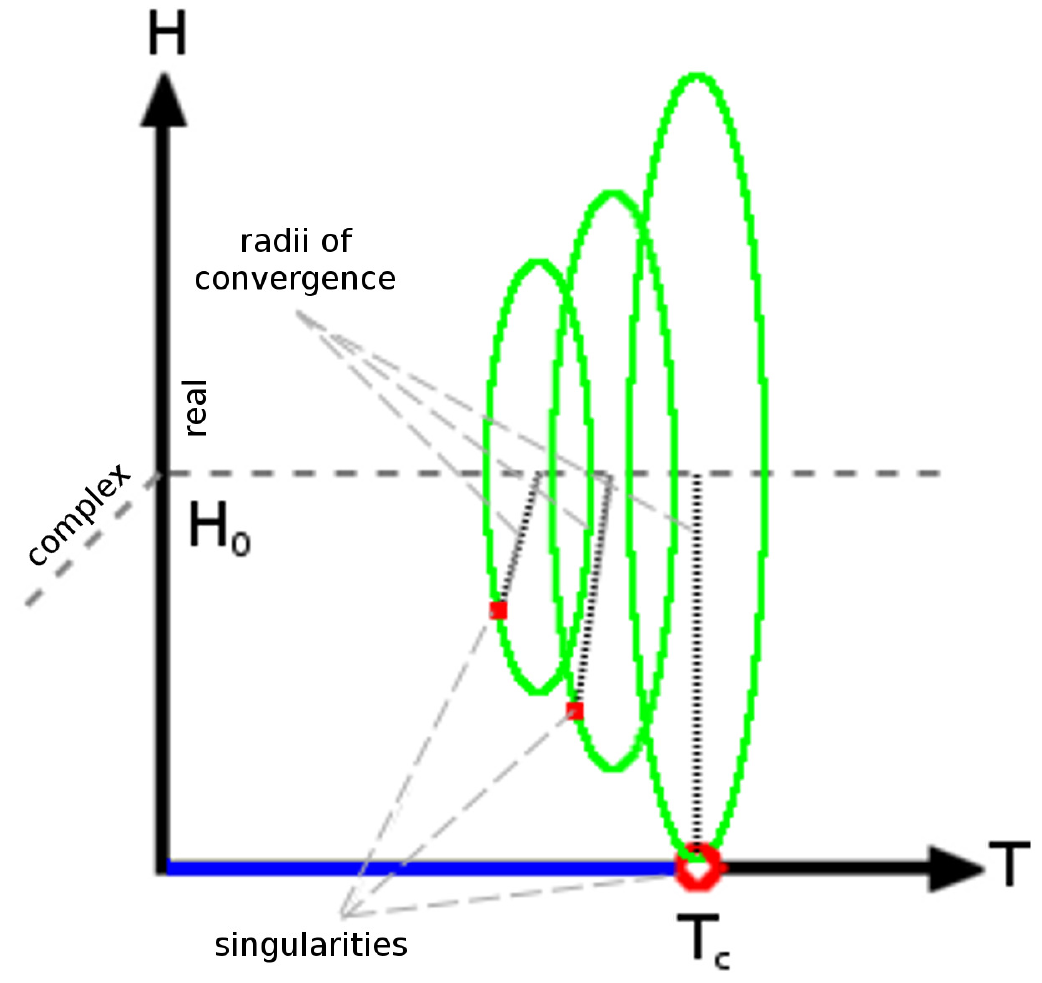}}
\subfigure[$~$QCD]
{\includegraphics[scale=0.7]{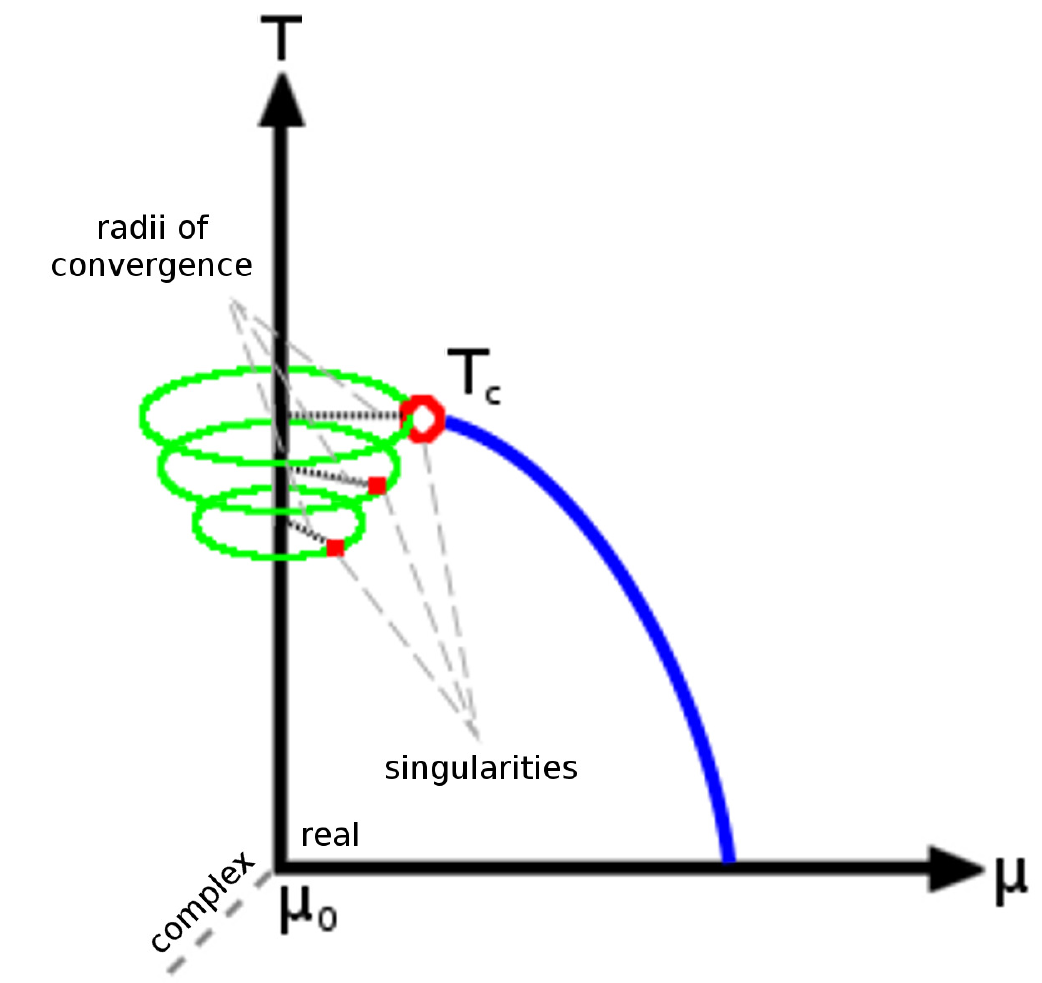}}
\caption{\label{fig:Us_Them}Visualization of the series expansion method
described in \cite{Gavai} and this work. Circles of radius
$r_\infty$ about any given expansion point are depicted in green. Domains
of convergence may not necessarily be finite sized for general
$T\neq T_c$ (i.e.
$r_\infty(T\neq T_c) = \infty$); however, when $r_\infty(T)$ is finite, the
corresponding singular point must be complex. Except, of course, when $T =
T_c$, at which point the singular point must be strictly real.}
\end{figure}

Looking at Fig.~\ref{fig:Us_Them}, the series expansion method in QCD
amounts to walking along the $\mu=0$ axis (where there is no sign
problem) and computing the Taylor coefficients of the quark number
susceptibility (QNS) at more or less evenly spaced intervals of $T$. It
is from these coefficients that one tries to learn about what takes
place at non-zero $\mu$. The direct analogy, in the Ising Model, is to
walk along some curve on Fig. \ref{fig:IsingPhaseDiagram}, and compute a
similar set of Taylor coefficients (again at roughly even
intervals). However, since there is no sign problem anywhere on
Fig. \ref{fig:IsingPhaseDiagram}, we are free to pick a convenient shape
for this curve. For the sake of simplicity, we will consider the problem
of locating the critical point based on correlation functions measured
along a line parallel to the $H=0$ axis, as depicted in
Fig. \ref{fig:Us_Them}.

Having computed the Taylor coefficients of the Ising Model free energy
$F$ up to some given order, we will then analyze the properties of this
truncated Taylor series in an attempt to locate the critical
point. Specifically, since the magnetic susceptibility diverges at the
critical point, we are looking for a non-analytic point of $F$. For the
Ising Model, this is a very convoluted way of locating the
critical point:  but that is the point.  We want to test how the
extrapolation method works in an example where we can also directly
locate the critical endpoint, to check our work.

The point is to mimic what has been done or at least is presently
\textit{doable} for QCD. Due to the ease with which one can perform
statistical simulations of the Ising Model we are theoretically able to
compute the Taylor coefficients needed for this analysis up to some very
high order, so we are not restricted in this regard. In QCD, this is
definitely not the case. Hence, we really only want to go to high enough
order to at least address the issue of what is needed to learn
\textit{anything}.

\section{Taylor Series about $H=H_0$ vs. Taylor Series about $\mu = 0$}

For an overview of the Taylor series method applied to QCD the reader is
referred
to \cite{Gavai}. Furthermore, a detailed overview of the Ising Model is given
in \cite{Domb}. We will now show how this translates over to the Ising Model,
which, in cartoon form, is illustrated in Fig. \ref{fig:Us_Them}. Assuming that
the QCD free energy is extensive, the
thermodynamic pressure is given by

\begin{equation}
 P_{\text{QCD}} = \frac{T}{V}\log \mathcal{Z}_{\text{QCD}}
\end{equation}
which can be expanded in a dual Taylor series in the up/down quark
chemical potentials about $\mu_u = \mu_d =0$
\begin{equation}
 \label{eq:QCD_Pres_Exp}
 P_{\text{QCD}}(T,\mu_u,\mu_d)= P_{\text{QCD}}(T,0,0) +
\sum_{n_u,n_d}\chi_{n_u,n_d}\frac{\mu_u^{n_u}}{n_u!}\frac{\mu_d^{n_d}}{n_d!}.
\end{equation} 
This series can be reformulated in terms of baryon 
($\mu_b = \frac{3}{2}(\mu_u + \mu_d)$) and isovector
($\mu_i = \frac{1}{4}(\mu_u - \mu_d)$) chemical potentials.  Fixing
$\mu_u = \mu_d$ and defining $\mu=\mu_b/3$ leaves a simplified
Taylor series in $\mu^2 = \mu_u^2 = \mu_d^2$.

The various $\chi_{i,j}$'s have expressions in
terms of the expectation values of products of $M_f$, $M_f^{-1}$
and its derivatives. Moreover, since the origin of the Taylor series is
$\mu = 0$, all of these computations are performed with
$\text{Det} M_f$ real and non-negative.

To construct the 3D Ising Model analogue, we begin by defining an $N \times N
\times N$ lattice $\Sigma$ as a set of individual
spins $\sigma_i$ on a 3D simple cubic lattice with volume $V=N^3$ and lattice
spacing $a=1$
\begin{equation}
 \Sigma = \{ \sigma_1, ..., \sigma_V\}\,.
\end{equation}
These spins interact via the Hamiltonian
\begin{equation}
 \mathcal{H} = - H \sum_i \sigma_i - J \sum_{\langle i,j \rangle} \sigma_i
\sigma_j 
\end{equation}
where the second summation is over all nearest neighbor pairs, and $J>0$ is
ferromagnetic. Given $\mathcal{H}$, an expression for the partition
function then follows. Defining the total spin $\sigma$
\begin{equation}
 \sigma = \frac{1}{V}\sum \sigma_i,
\end{equation}
we have
\begin{equation}
 \mathcal{Z} = \text{tr} \{ e^{-\beta\mathcal{H}} \} 
  = \sum_\Sigma e^{\beta H V
\sigma} \prod_{\langle i,j \rangle}e^{\beta J \sigma_i\sigma_j}
\end{equation}
where, as usual, $\beta = 1/T$. Hence, the free energy per
unit volume is
\begin{equation}
 F = -\frac{T}{V} \log \mathcal{Z}
\end{equation}
from which we obtain the thermodynamic variables
\begin{eqnarray}
 M &=& -\Big ( \frac{\partial F}{\partial H}\Big)_T\\
 \chi &=& -\Big ( \frac{\partial^2 F}{\partial H^2}\Big )_T
\end{eqnarray}
where $M$ and $\chi$ are respectively the magnetization and magnetic
susceptibility. A Taylor series for $F$ is obtained by simply noting that
$\mathcal{Z}$ is a moment generating function, and hence $\log \mathcal{Z}$
is a cumulant generating function \cite{Lauritzen}. Expressions for the various
cumulants
$\kappa_n$ in terms of the moments $\langle \sigma^n \rangle$ can be obtained
from the generating function
\begin{equation}
 \kappa_n = \frac{d^n}{d t^n} \log \langle e^{t\sigma} \rangle \Big \vert_{t=0} 
\end{equation}
and, for instance,
\begin{eqnarray}
 \kappa_1 &=& \langle \sigma \rangle \nonumber\\
 \kappa_2 &=& \langle\sigma^2 \rangle - \langle \sigma \rangle^2 \nonumber\\
 \kappa_3 &=& \langle \sigma^3 \rangle - 3 \langle \sigma^2 \rangle \langle \sigma \rangle + 2\langle \sigma \rangle^3 \nonumber\\
 \kappa_4 &=& ... \label{eq:cumulants}
\end{eqnarray}
Thus, the Taylor series for $F$ is given by
\begin{equation}
\label{eq:Fkappa}
 F(H;H_0,T) = -\frac{T}{V}\log \mathcal{Z}(H_0,T) -
\frac{T}{V}\sum_{n=1}^\infty \frac{V^{n}}{n!} \kappa_n \Big
( \frac{H - H_0}{T}\Big )^n
\end{equation} 
which is an expansion in $H$ at fixed $T$.

In summary, the analysis for QCD and the Ising Model amounts to extrapolating
the
behavior of the higher order $\chi_{i,j}$'s and the $\kappa_i$'s from those
that can be calculated. If a singularity is suspect,
its distance from the expansion point is equal to the series' radius of
convergence. For an arbitrary function $f(x)$ expanded in a series 
\begin{equation}
 f(x) = \sum_{n=0}^{\infty} f_n x^n
\end{equation}
the radius of convergence $r_\infty$ is given formally by 
\begin{equation}
\label{eq:rinfty}
r_\infty = \frac{1}{\limsup_{n\rightarrow \infty} \sqrt[n]{|f_n|}} .
\end{equation}
In our case, we may infer the radius of convergence from the asymptotic behaviour of
\begin{equation}
\label{eq:r_n}
r_n = \Big \vert \frac{f_n}{f_{n+1}}\Big \vert .
\end{equation}
Alternatively, we may also consider the
$n\rightarrow\infty$ limit of 
\begin{equation}
\tilde{r}_{n} = \Big \vert \frac{f_{2n}}{f_{2n+2}}\Big \vert^{\frac{1}{2}} 
\end{equation}
or
\begin{equation}
 \rho_n = \Big \vert \frac{f_{0}}{f_{2n}} \Big \vert ^{\frac{1}{2n}}.
\end{equation}

All of these definitions are effectively equivalent in that they all yield the
same $n\rightarrow \infty$ limit. However, $\tilde{r}_{n}$
and $\rho_n$ are of relevance to \cite{Gavai} since the Taylor series that they
analyze only has even
powers of $\mu$. This is, however, a minor technicality. In both models, the
asymptotic behavior of $r_n$ (or $\tilde{r}_{n}$ or $\rho_n$) is inferred from
the series' few
lowest order Taylor coefficients. Specifically, the highest order that is
achieved for QCD is $n=8$. So that we remain on roughly the same footing, we
will thus
restrict our analysis to $n\leq 8$ coefficients.

Monte Carlo studies \cite{Talapov} with lattices up to $N = 256$ have
revealed the Ising critical point to be at $J / T_c = 0.2216544(6)$ (i.e. $T_c/J
\approx 4.51$), and
of course we already know that $H_c = 0$. What we
now want to figure out
is whether or not we can infer the location of this critical
point by carrying out the expansion described above.
 
\section{Results\label{sec:Results}}

Monte Carlo simulations were performed on 3D simple cubic lattices using
the Cluster Algorithm \cite{Krauth}.  As mentioned in the introduction,
we are restricted on the phase diagram to lines of constant background
magnetic field strength, denoted by $H_0$. We specifically choose to
work along the lines of $H_0 = 0.1J$ and $H_0 = 0.01J$, since at these
distances we are able to obtain good statistics, while remaining in the
critical region (when near $T_c$).  At distances too close to the
critical point the correlation length grows rapidly, hence much larger
volumes would require consideration. Whereas, if one is too far from the
critical point ($H_0=J$, say), critical behavior is not observed.
For the values of $H_0$ that we considered, it was sufficient to consider
sizes in the range $4\leq N \leq 24$.

Now that we have chosen the values of $H_0$ around which we will base
our analysis, we need to determine the minimal lattice sizes (for each
of our the two different values of $H_0$) which sufficiently
converge to the thermodynamic limit. Due to the divergence of the
correlation length at the critical point, one would expect that
obtaining reasonable data for $H_0 = 0.01J$ requires a larger (perhaps
significantly) lattice than for $H_0 = 0.1J$.

Such convergence is best observed qualitatively by plotting $\kappa_4$
as a function of $T$ over a range of volumes (shown in
Fig.~\ref{fig:H01VolDep} for $H_0 = 0.1J$ and Fig. \ref{fig:H001VolDep}
for $H_0 = 0.01J$). These graphs suggest that it is sufficient to work
with $N=8$ and $N=20$ for $H_0 = 0.1J$ and $H_0 = 0.01J$ respectively,
based on qualitative observation that the thermodynamic ``limit'' is
achieved whenever any two subsequent graphs overlap.

More to the point, the thermodynamic limit is effectively reached when
the volume dependence of higher order cumulants becomes washed away in
statistical uncertainty. Said differently, it becomes impossible to
accurately distinguish volume dependence from noise. This effect is
shown in Figs. \ref{fig:K8H01VolDep} and \ref{fig:K8H001VolDep} for
$\kappa_8$, where it is made very clear for $H_0 = 0.01J$. The
increasing uncertainty eventually renders the analysis inconclusive at
sufficiently large volumes.

Based on Figs. \ref{fig:H01VolDep} and \ref{fig:K8H01VolDep} it may seem
a bit overzealous to regard $N=8$ as the thermodynamic limit. However,
at the expense of relatively small finite-size volume effects, it turns
out to be very statistically advantageous to work with $N=8$ over say
$N=9$; this is further discussed in the next section. It should be noted
that measurements of the correlation length along our values of $H_0$
(see Appendix A) further validate the choices of $N=8$ and $N=20$, since
it turns out that these lattices are many correlation lengths across.

Henceforth, if we are discussing $H_0=0.1J$ results it should be assumed
that the data originated from an $N=8$ lattice and likewise $N=20$ for
$H_0=0.01J$, unless it is stated otherwise. Moreover, it should be
assumed that any measured quantity is obtained from averaging over
$\mathcal{N} = 10^7$ uncorrelated samples, with estimates of uncertainty
calculated using the Jackknife Method \cite{Chernick}. 

The results of our analysis are plotted over
Figs. \ref{fig:Magnetization} to
\ref{fig:chi45}. Fig. \ref{fig:Magnetization} illustrates the crossover
from the magnetized to the unordered phase ($\kappa_1$ is simply the
magnetization per unit volume). The Taylor coefficients (higher order
$\kappa_i$'s, normalized to unity) are shown in
Fig. \ref{fig:KappaAll}. The Taylor series constructed out of these
coefficients are depicted in Figs. \ref{fig:chi6} and
\ref{fig:chi45}. Finally, $r_n$ (from which we extrapolate radii of
convergence) are shown in Figs. \ref{fig:Rn} to \ref{fig:H001Rn}.

\section{Analysis and Discussion}

Now that the data has been presented, the first thing that we should try to do
is infer (guess) the location of $T_c$. Subsequently, we will then
calculate $H_c$ and specify the location of the critical point (assuming that
everything goes according to plan).

Monte Carlo simulations of the Ising Model are drastically simpler than those
of QCD. That being said, the ``restricted'' extent of our data set is not due to
any sort of numerical limitations. If we truly desired we could easily a) obtain
data from larger
sample sizes, b) sample the temperature axis more frequently and c) obtain even
higher order Taylor coefficients. Rather, as discussed in the introduction, we
are on purpose trying to somewhat mimic the limitations facing
any realistic study of QCD. 

In what follows, we will try to apply to same logic in locating the Ising critical point as that which has been used for QCD. Therefore, a bit of a synopsis of the analysis in \cite{Gavai} would be in order. The authors focus mainly on the Taylor expansions of $\frac{\partial^2 P_{\text{QCD}}}{\partial \mu_u^2}$ and $\frac{\partial^2 P_{\text{QCD}}}{\partial \mu_u \partial \mu_d}$ up to sixth order\footnote{Note that even when $m = m_u = m_d$ and $\mu=\mu_u=\mu_d$, $\chi_{2,0}$ does not necessarily equal $\chi_{1,1}$ (flavour symmetry only requires $\chi_{2,0} = \chi_{0,2}$). Simply, $\chi_{1,1} \sim \text{Det}'\text{Det}'$ while $\chi_{2,0} \sim \text{Det}''\text{Det}$.}. By definition, expanding in $\mu = \mu_B/3$ and $\mu_I = 0$,
\begin{equation}
\frac{\partial^2}{\partial \mu_u^2} P_{\text{QCD}}(T,\mu) = \chi^0_{B} +  \chi^2_{B} \mu^2 + \chi^4_{B} \mu^4 + \chi^6_{B} \mu^6 + ...,
\end{equation}
where $\chi^0_{B} = \chi_{2,0}$ and, for instance $\chi^4_{B} = (\chi_{6,0} + 4\chi_{5,1} + 7\chi_{4,2} + 4\chi_{3,3})/4!$ with similar expressions for the other $\chi_B$'s. The critical region is highlighted by the formation of sharp peaks in the $\chi_{B}$'s, which are observed in the vicinity of $T_{c/o}$. Furthermore, at $T < 0.95 T_{c/o}$, $\chi^6_{B}$ changes sign. By plotting the Taylor series of $\frac{\partial^2 P_{\text{QCD}}}{\partial \mu_u^2}$, criticality is most evident at $T = 0.95 T_{c/o}$, since this temperature corresponds to the most rapidly diverging graph. Subsequently, the distance to the CEP if inferred from an extrapolation of the series' radius of convergence. 

Hence, in our case, the analysis will proceed in the following order, 
\begin{itemize}
\item locate the critical region
\item observe the range over which the Ising coefficients exhibit the correct alternating sign behavior indicative of a divergence along the negative $H$ axis 
\item estimate the exact location of $T_c$ by plotting $\chi$'s Taylor series 
\item finally, determine the location of the critical point from our extrapolations of $r_n$.
\end{itemize} 

Since we are not on the $H = 0$ axis, nor are we in the infinite volume
limit, the curves in Fig. \ref{fig:KappaAll} depict a crossover rather than a
phase transition. By observing the range over which $\kappa_8$ exhibits
non-trivial behavior, we can safely say that $T_c / J$ is bracketed between
$[4.25, 5.00]$ for $H_0=0.1J$ and $[4.45,4.65]$ for $H_0=0.01J$.

Continuing on, the singular point that
we are looking for lives on the negative $H$ axis of our Taylor series. In
other words, it is a requirement for our Taylor coefficients to alternate in
sign order by order. Looking back at Fig. \ref{fig:KappaAll}, we see that at
$T=4.80J$ and $T=4.60J$ for $H_0=0.1J$ and $H_0=0.01J$ respectively the
$7^{\text{th}}$
order coefficient changes sign, thus breaking the pattern. From this
observation we can refine our upper bounds on $T_c$.

At the order that we are working with, there is a whole range (rather than a
single point) in
$T$ where the coefficients are observed to alternate in sign. Hence, this property alone is not sufficient to specify $T_c$. If we were to plot the
full Taylor series for the magnetic susceptibility $\chi$, we would observe a
genuine singularity at the critical point $T_c$ (assuming that we are in the
thermodynamic limit). Aside from being at finite volume, we do not have the
full series, so the best that we can do is plot the partial sum of the first 6 terms (we
only have 6 rather than 8 terms since we wish to plot $\chi$ rather than
the free energy $F$, analogous to the situation in QCD).

The rate at which this polynomial diverges is directly attributable to the
values of the Taylor coefficients, {\it i.e.\ } the temperature at which
this polynomial diverges most rapidly should give a somewhat clearer
indication of the location of $T_c$. The partial sum of $\chi$'s Taylor
series up to order $n$ is given by
\begin{equation}
\label{eq:chi(n)}
 \chi^{(n)} = \Big (\frac{V}{T} \Big) \Big (\kappa_2 + \sum_{i = 1}^{n}
\frac{V^{n} \kappa_{n+2}}{n!} \frac{(H - H_0)^n}{T^n}\Big ).
\end{equation}

$\chi^{(6)}$ is plotted in Fig. \ref{fig:chi6}. Uncertainties, when depicted,
are due to the standard error of the highest order Taylor coefficient; in this
case that would be $\kappa_8$. When $H_0=0.1J$, the $T=4.50J$, $T=4.55J$ and
$T=4.60J$ graphs, aside from being
indistinguishable, diverge most rapidly. From the plots we expect that
$4.50 \lesssim T_c / J \lesssim 4.60$. Furthermore, the $H_0=0.01J$ plot suggests that $T_c /
J \simeq 4.50$.

Now that we are led to believe that $T_c / J = 4.55 \pm 0.05$ (or least
that $T_c$ is in the vicinity of $4.55J$), we will try to
locate $H_c$. In the infinite volume limit, 
\begin{equation}\label{eq:Hc}
H_c = H_0 - T_c r_\infty(T_c), 
\end{equation}
where $r_\infty(T)$ is the radius of convergence as defined earlier, which in general is temperature dependent, and most importantly $r_\infty(T_c)$ is finite due to non-analyticity at the critical point. However, for any finite volume, the partition function is analytic everywhere on the phase diagram, hence the radii of convergence that we measure should \textit{always} diverge when $n\rightarrow \infty$. So, in fact, when we say that we are extrapolating $r_n$ to $r_\infty$ (i.e. taking the limit defined by Eq. \ref{eq:rinfty}), what we really mean is that we expect to observe a range in values of $n$ where $r_n$ forms a plateau. The value of $r_n$ along this plateau is used to determine $H_c$ via Eq. \ref{eq:Hc}.

In Fig. \ref{fig:Rn}, we observe exactly this sort of behavior. $r_n(T)$
forms a plateau for some (but not all) values of $T$ in
the vicinity of $T_c$ (shown in greater detail in Fig. \ref{fig:H001Rn}). With $H_0=0.01J$, a plateau forms when $T=4.50J$, and it continues all the way up to $n=7$. When we go to $T=4.55J$, we see that by $n=7$ the plateau has started to tilt upwards. The $H=0.1J$ graphs exhibit similar behavior, see Fig. \ref{fig:H01Rn}.

For both values of $H_0$, $r_n$ has the smallest error bars when $T=4.50J$; furthermore, the plateaus that form at this temperature extend all the way up to $n=7$.
Hence, our most reliable estimate of $H_c$ can be obtained from the $T=4.50J$
graphs. Within error bars, $-0.005 < H_c/J < 0.015$ from $H_0 = 0.1J$ and
$-0.002 < H_c/J < 0.002$ from $H_0 = 0.01J$.

Thus, from the analysis centered about $H_0=0.1J$, we can conclude that $4.50 \lesssim T_c /
J \lesssim 4.60$ and $-0.005 \lesssim H_c/J \lesssim 0.015$, a finding which is
in perfect agreement with the known location of the Ising critical point. Therefore, we have observed that this method is able to effectively locate the Ising Model
critical point.

In working with the Ising model, we were not limited numerically, hence we
carried out this analysis to $8^{\text{th}}$ order in the cumulant
expansion without much difficulty. In doing so, we were able to make a
few observations that may be of relevance to a study revolving around
QCD.

Foremost, our best data came from the $N=8$ and $N=20$ lattices, which
were the minimum sizes that could be considered ``thermodynamic.'' As we
discussed, the subtraction of disconnected pieces from high order
cumulants leads to a very rapid deterioration with volume in the
statistics.  Therefore one must make a careful balancing act between
large volumes to control finite-volume effects, and the loss of
statistics in high-order cumulants with large volume.

It is also interesting to see how the effectiveness of the method
changes as one is limited to lower orders in the Taylor series. Plots of
$\chi^{(4)}$
and $\chi^{(5)}$ are shown in Fig. \ref{fig:chi45}. When limited to
$6^{\text{th}}$ order, it would appear that $ 4.60 \lesssim T_c/J
\lesssim 4.65$. This is a slight overestimate, but the real issue here
is that it becomes impossible to determine $H_c$. In Figs. \ref{fig:Rn},
\ref{fig:H01Rn} and \ref{fig:H001Rn}, we see that
$r_n$ only forms a plateau for $n\geq5$. At $8^{\text{th}}$ order, we
extrapolate $r_n$ solely from $r_5$, $r_6$ and $r_7$. However, when restricted
to $6^{\text{th}}$ order, no such extrapolation is possible, since you
would only know $r_5$.  Recall that, since the expansion in QCD in terms
of the chemical potential only contains even powers in $\mu$, the 4'th
order cumulant in the Ising model is equivalent to the $\mu^8$ term in
QCD.

\section{Conclusion}

We have tested the prospects for the Taylor Series Technique for
locating a critical endpoint in a phase space which cannot be explored
directly.  By finding the cumulants in terms of the unexploreable
direction, one can extrapolate the radius of convergence and type of
convergence to find a critical temperature and other parameter (chemical
potential in QCD).  In the ``toy'' case where we confine ourselves to
the $H_0=0.1J$ line of the 3D Ising model phase
diagram, we calculated the Taylor coefficients in an expansion of the free
energy in $H$.
By analyzing the properties of these Taylor coefficients, as well as
extrapolating the radius of convergence of the series which they form, we were
led to conclude that the critical point lives in the range $4.50 \lesssim
T_c/J \lesssim 4.60$ and $-0.005 \lesssim H_c/J \lesssim 0.015$. This finding is in agreement with the
known result that $T_c / J \approx 4.51J$ and $H_c/J = 0$. 
Therefore, we find that the Taylor series method can be applied
successfully to find a critical endpoint outside of the range where
direct lattice studies are conducted. In this sense,
the findings of this analysis offer a positive outlook for the
applicability of the method to QCD.

However, we found that a successful extrapolation of the critical point
requires high orders in the cumulant expansion, as well as excellent
control of errors.  In the case we considered, an accurate determination
of the distance to the critical endpoint required the use of 8 terms in
the cumulant expansion; using only 4 terms (the number currently
available in the case of QCD) is not enough to determine the distance to
the critical point accurately.

\appendix

\section{Correlation Lengths}

\begin{figure}
\centering
\subfigure[$~H_0=0.1J$\label{fig:K8H01VolDep}]
{\includegraphics[scale=0.4,angle=270]{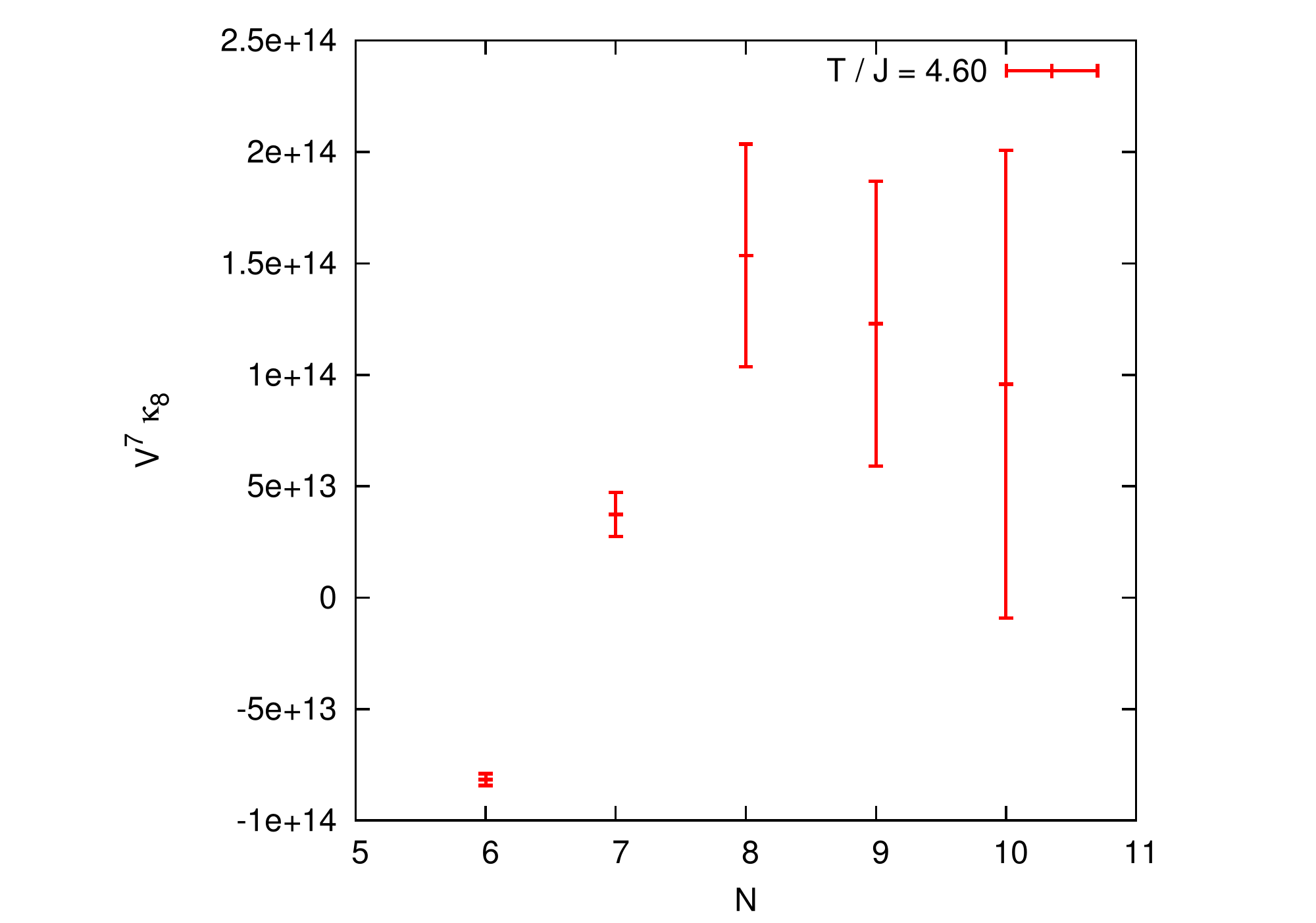}}
\subfigure[$~H_0=0.01J$\label{fig:K8H001VolDep}]
{\includegraphics[scale=0.4,angle=270]{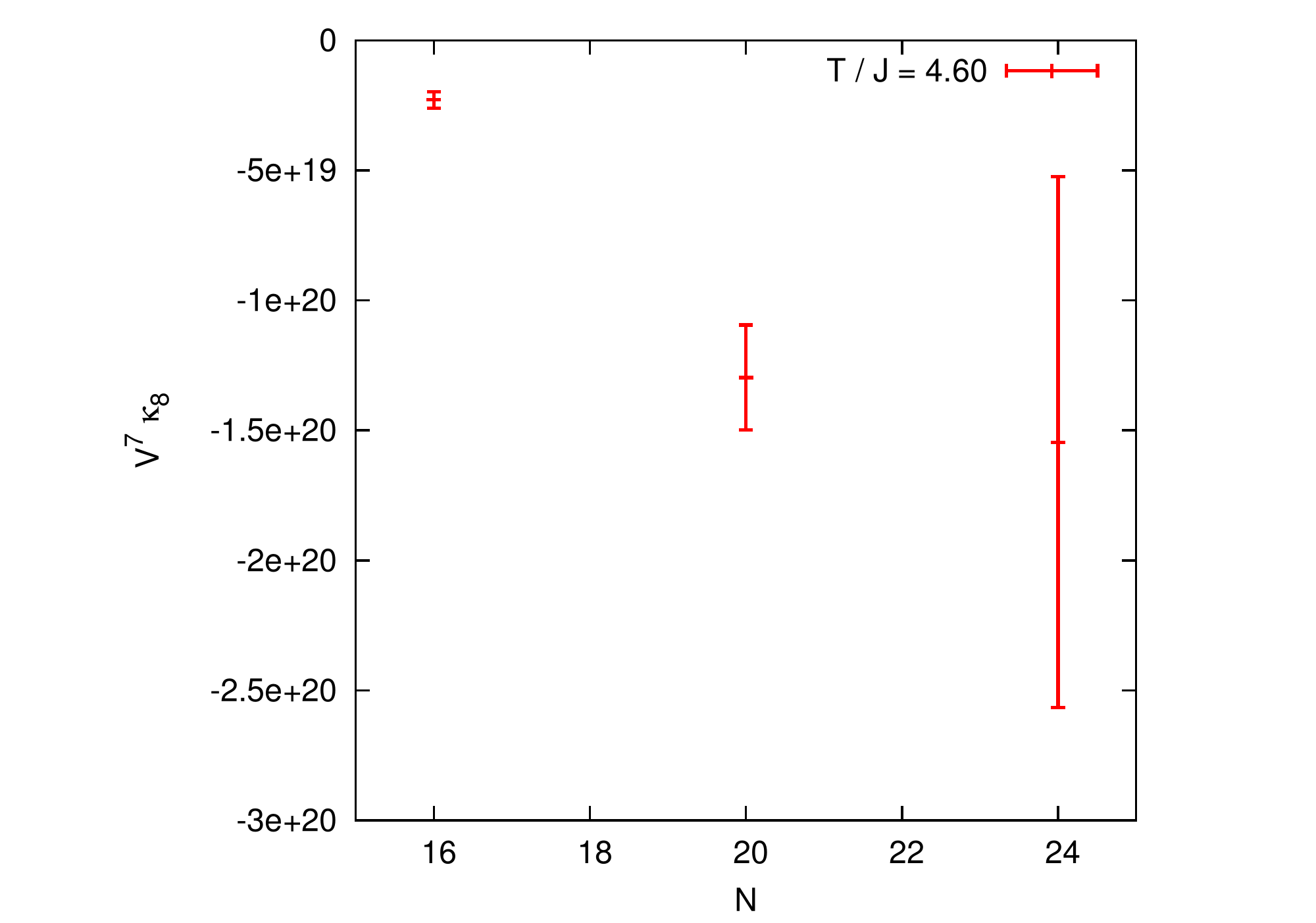}}
\caption{Volume dependence of $\kappa_8$ at $T = 4.60 J$ (note the volume normalization, the quantity plotted is actually $V^7\kappa_8$).}
\end{figure}

In Section \ref{sec:Results} we concluded that $N=8$ and $N=20$ lattices
sufficiently
approximate the thermodynamic limit when $H_0 = 0.1J$ and $H_0 = 0.01J$
respectively. This was based primarily on observations of the volume dependence of
$\kappa_8$ (see Figs. \ref{fig:K8H01VolDep} and \ref{fig:K8H001VolDep}).

Alternatively, we could have reached this conclusion by calculating correlation
lengths, denoted by $\xi$. For our purposes it was sufficient to qualitatively
observe the convergence of $\kappa_8$, however, it is still of interest to
examine the thermal dependence of $\xi$ for both background field strengths.
This will serve as a non-trivial check that it was sensible to base our findings
on $N=8$ and $N=20$ lattices, i.e. we should observe that $N\gg \xi$.

$\xi$ can be calculated by measuring the exponential fall-off of the
two-point function $G(r) = \langle \sigma (x) \sigma (0) \rangle$. The tail of $G(r)$ was
subsequently fitted to
\begin{equation}
\label{eq:Gtilde}
 \tilde G(r) = \frac{C e^{-\frac{r}{\xi}}}{r^{\text{D}-2+\eta}} + M^2
\end{equation}
since $\tilde G$ exhibits the correct asymptotic behaviour. $\eta$ was fixed at 0.025, in accordance with the results found in \cite{Gupta}.

\begin{figure}
\centering
\includegraphics[scale=0.7,angle=270]{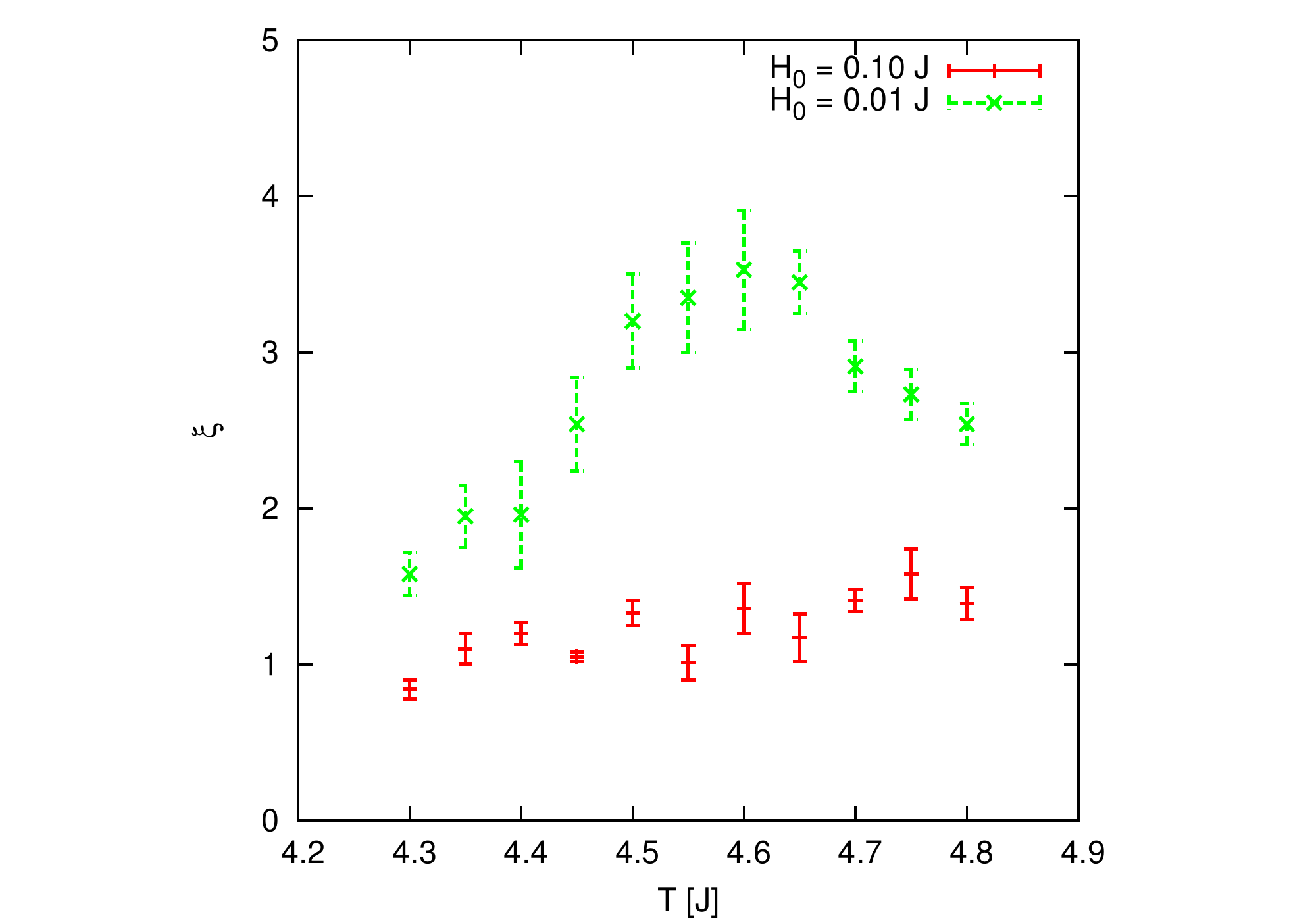}
\caption{\label{fig:Xi}Correlation length $\xi$, obtained by fitting
Eq. \ref{eq:Gtilde} to $\langle \sigma(r) \sigma(0) \rangle$, illustrating why
it was necessary to go to $N=20$ for $H_0=0.01J$.}
\end{figure}

Based on the results of this fit, $\xi$ as a function of temperature is plotted
in Fig. \ref{fig:Xi}. These graphs are somewhat noisy since measurements of
$G(r)$ were obtained from $\mathcal{N}=10^3$ uncorrelated lattices. However,
what we observe
is that $\xi \approx 1$ when $H_0=0.1J$, while $\xi$ jumps to near 4 when
$H_0=0.01J$, justifying the necessity to go to a larger volume.

\begin{figure}
\centering
\subfigure[$~H_0=0.1J$\label{fig:H01VolDep}]
{\includegraphics[scale=0.4,angle=270]{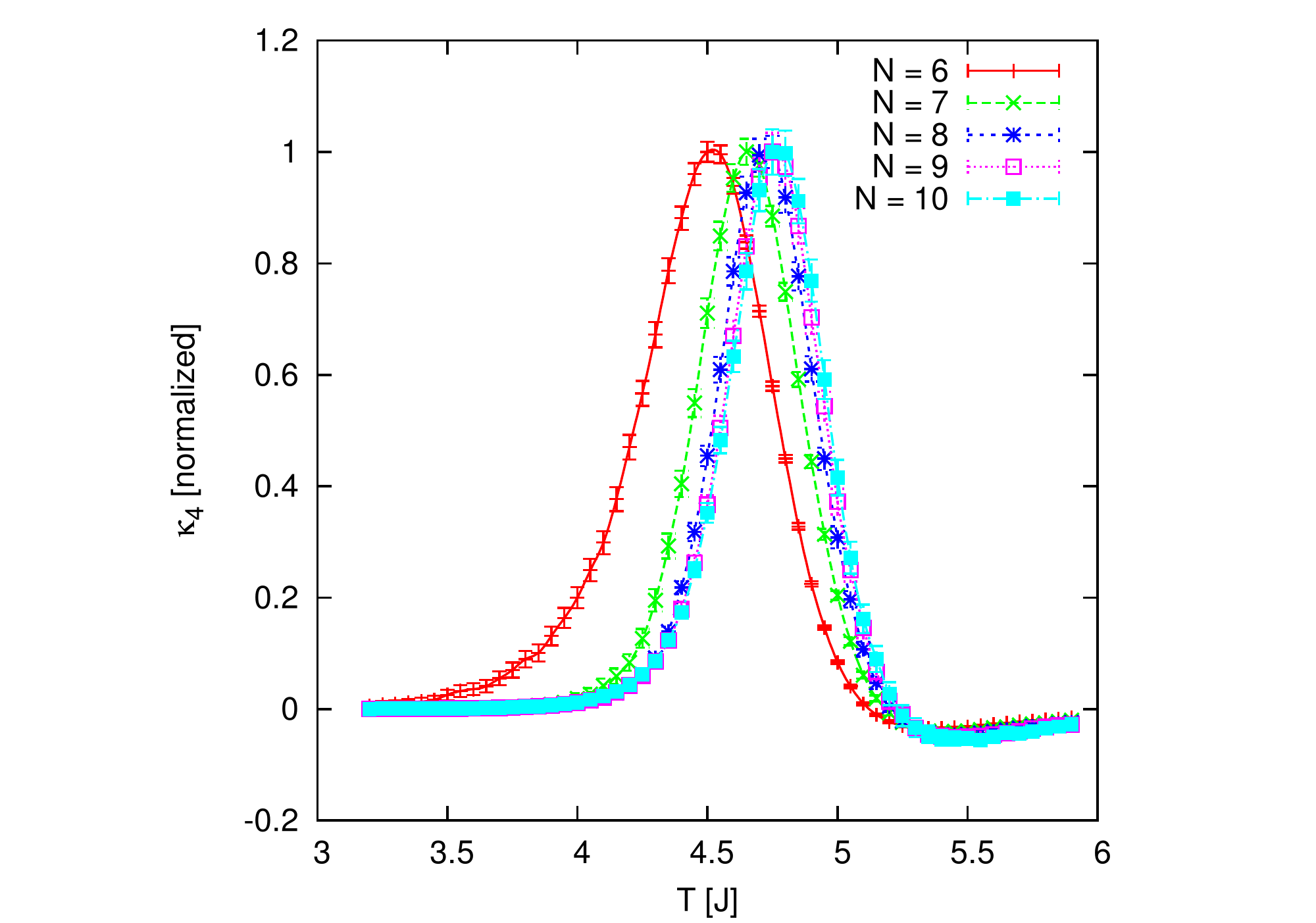}}
\subfigure[$~H_0=0.01J$\label{fig:H001VolDep}]
{\includegraphics[scale=0.4,angle=270]{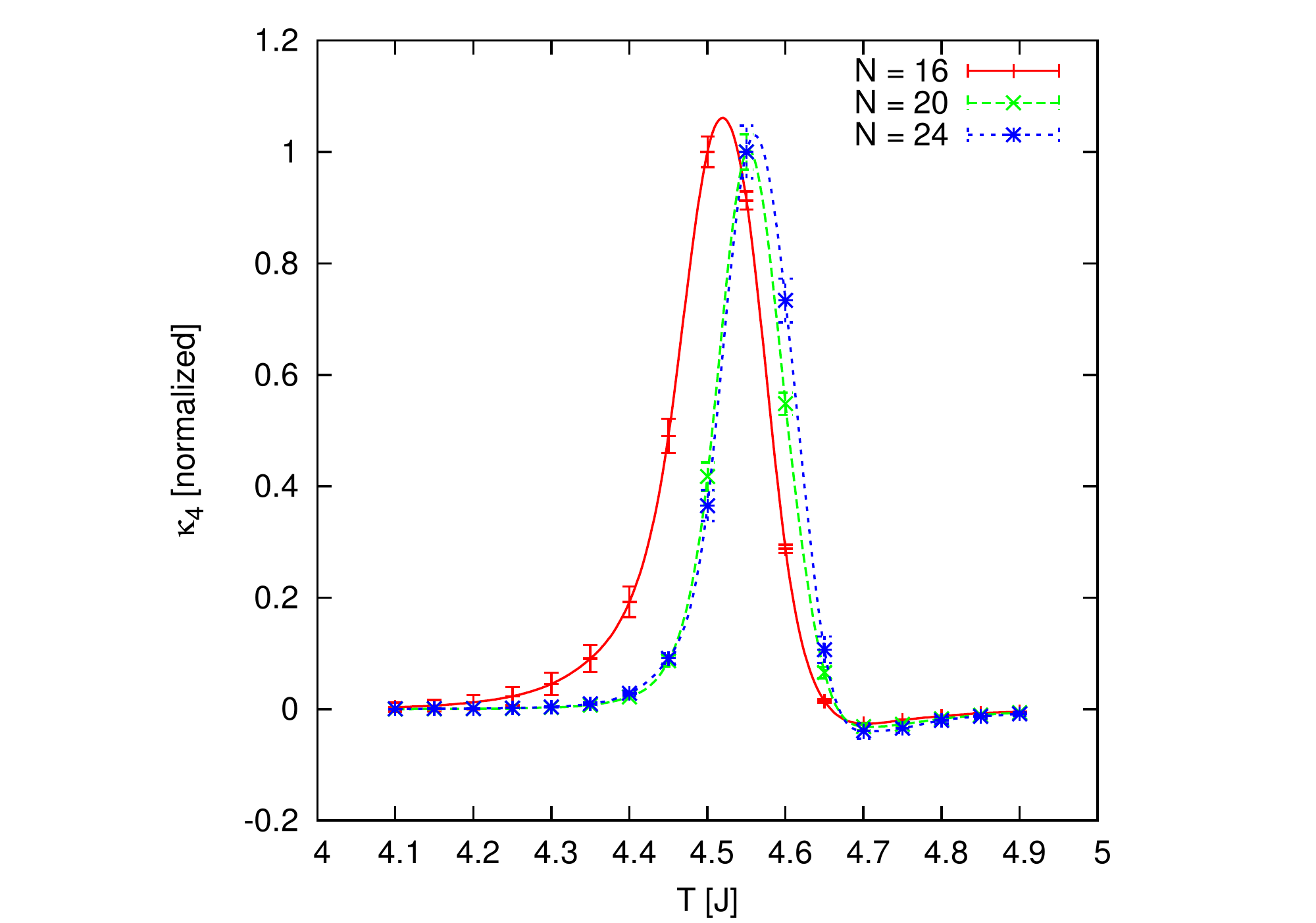}}
\caption{Volume dependence of $\kappa_4$ indicating convergence to the
thermodynamic limit.}
\end{figure}

\begin{figure}
\centering
\subfigure[$~H_0=0.1J$]
{\includegraphics[scale=0.4,angle=270]{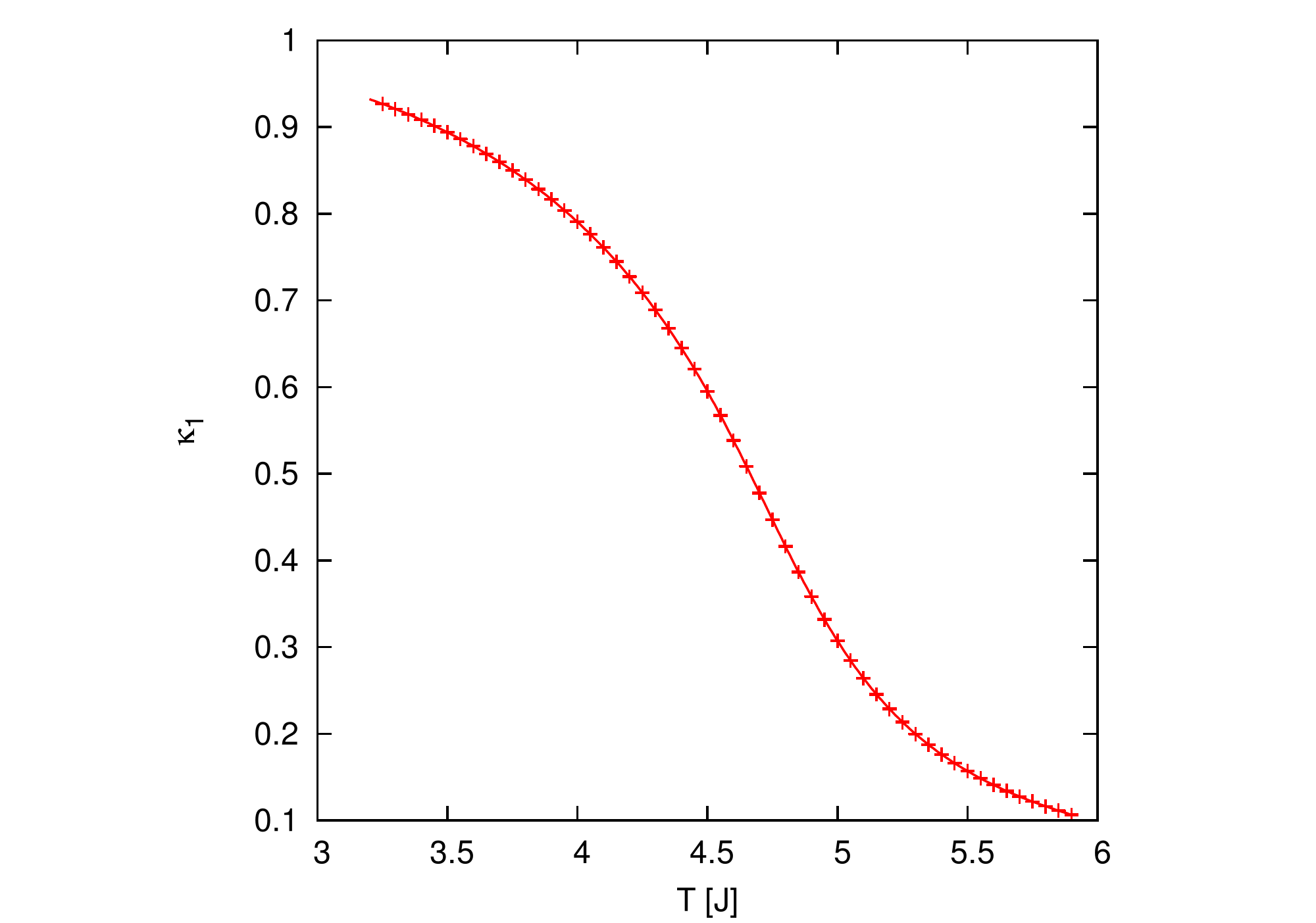}}
\subfigure[$~H_0=0.01J$]
{\includegraphics[scale=0.4,angle=270]{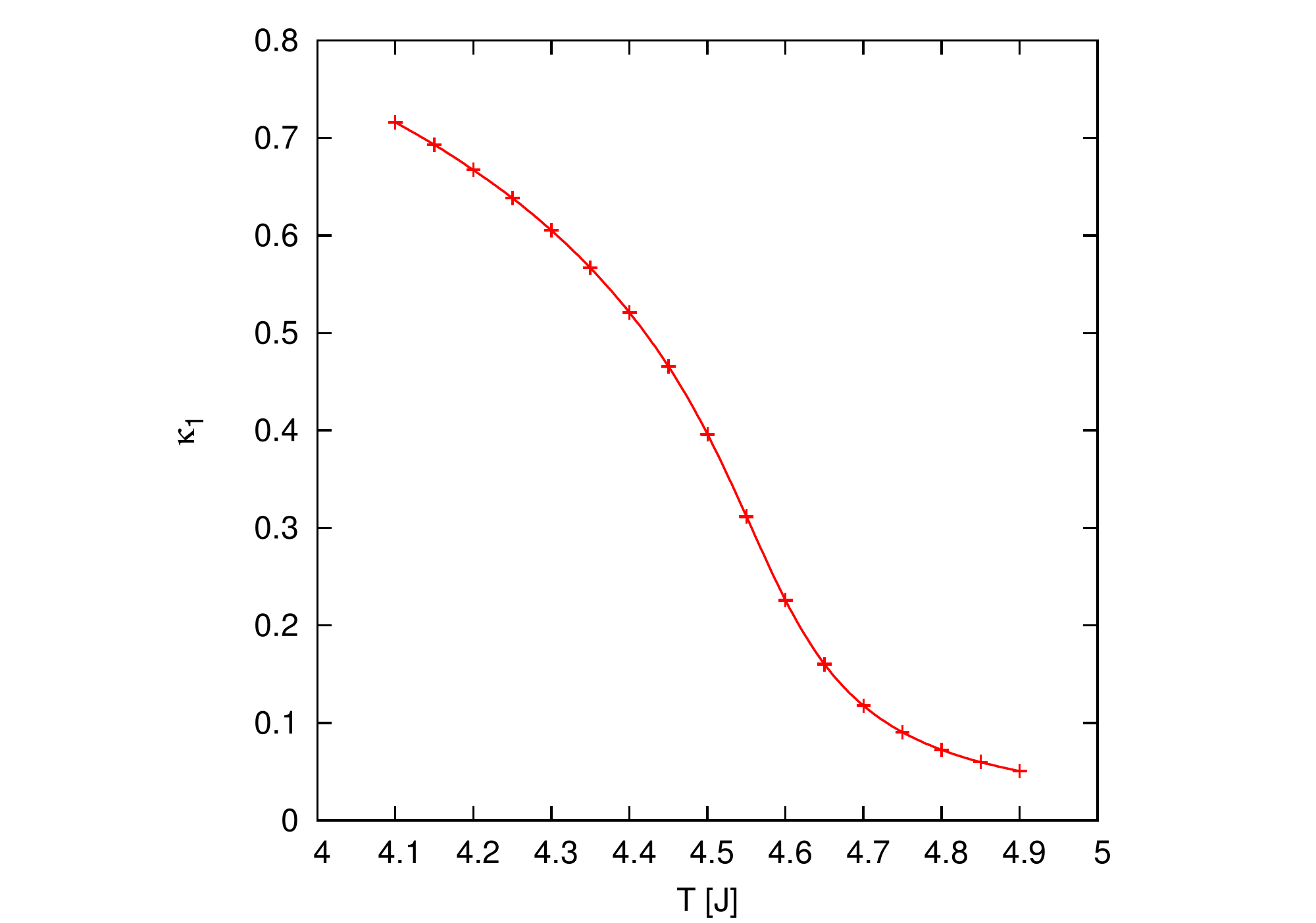}}
\caption{\label{fig:Magnetization}Temperature dependence of the magnetization
per unit volume, showing the crossover between the ordered / unordered phases.}
\end{figure}

\begin{figure}
\centering
\subfigure[$~H_0=0.1J$]
{\includegraphics[scale=0.4,angle=270]{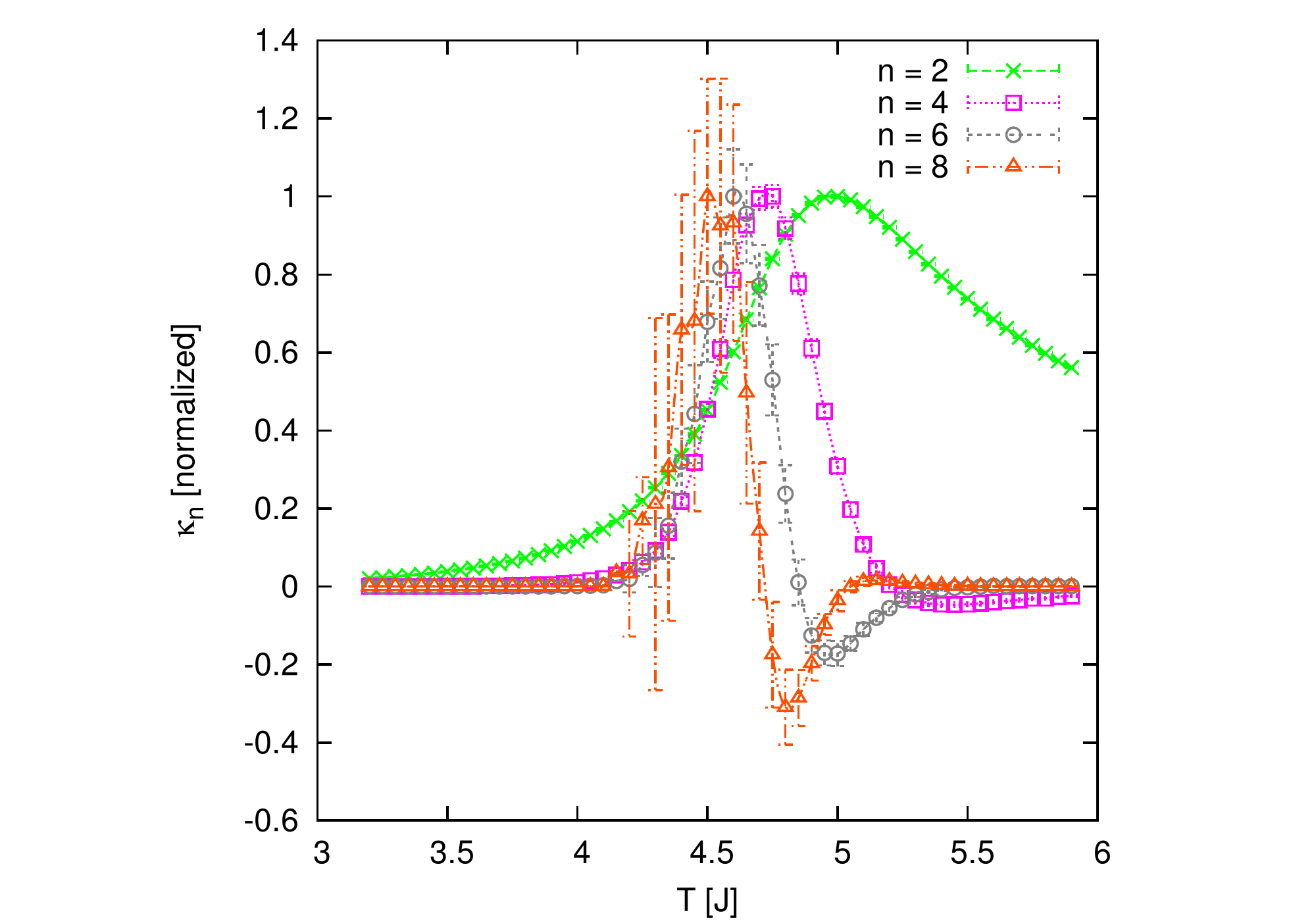}}
\subfigure[$~H_0=0.1J$]
{\includegraphics[scale=0.4,angle=270]{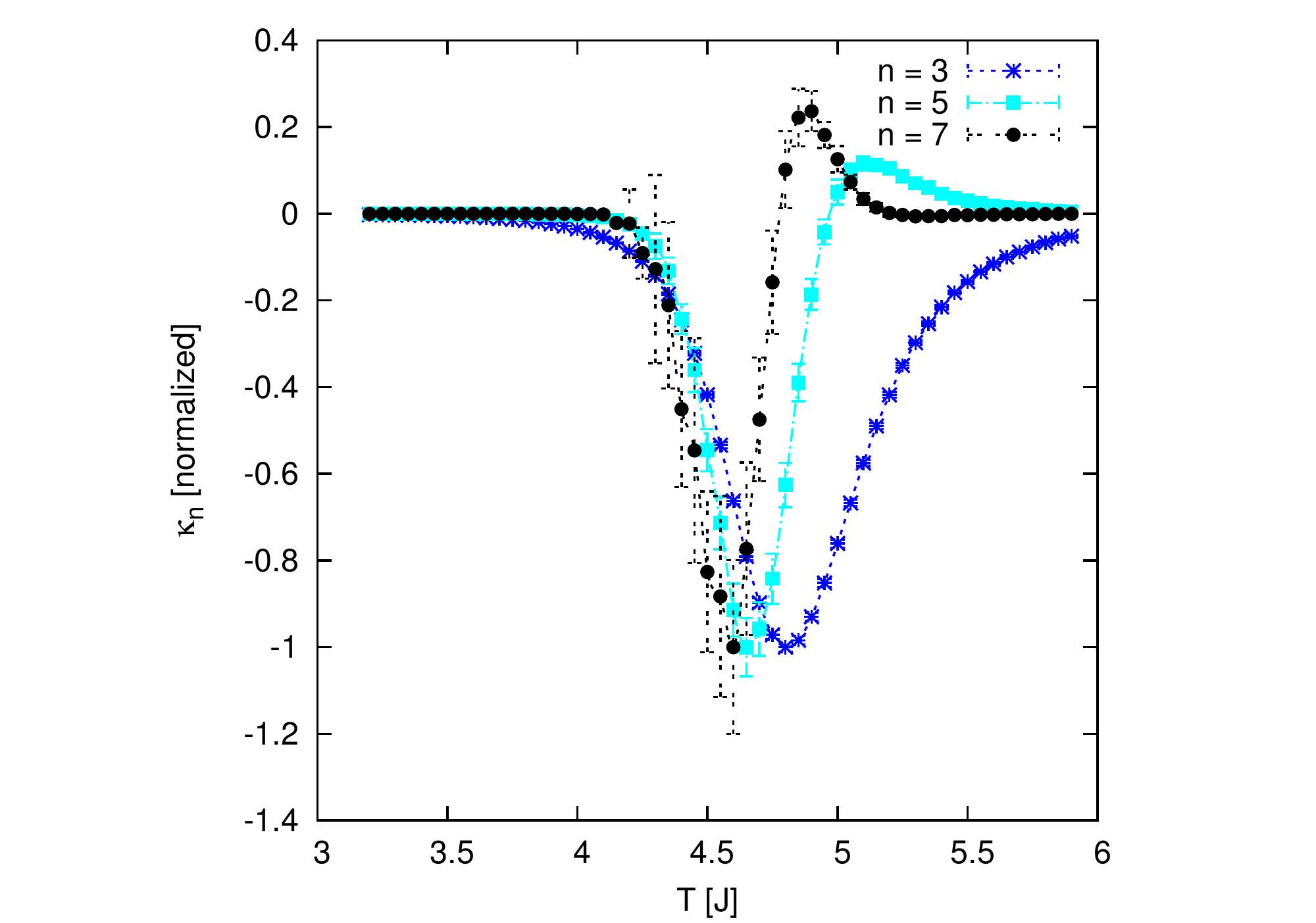}}\\
\subfigure[$~H_0=0.01J$]
{\includegraphics[scale=0.4,angle=270]{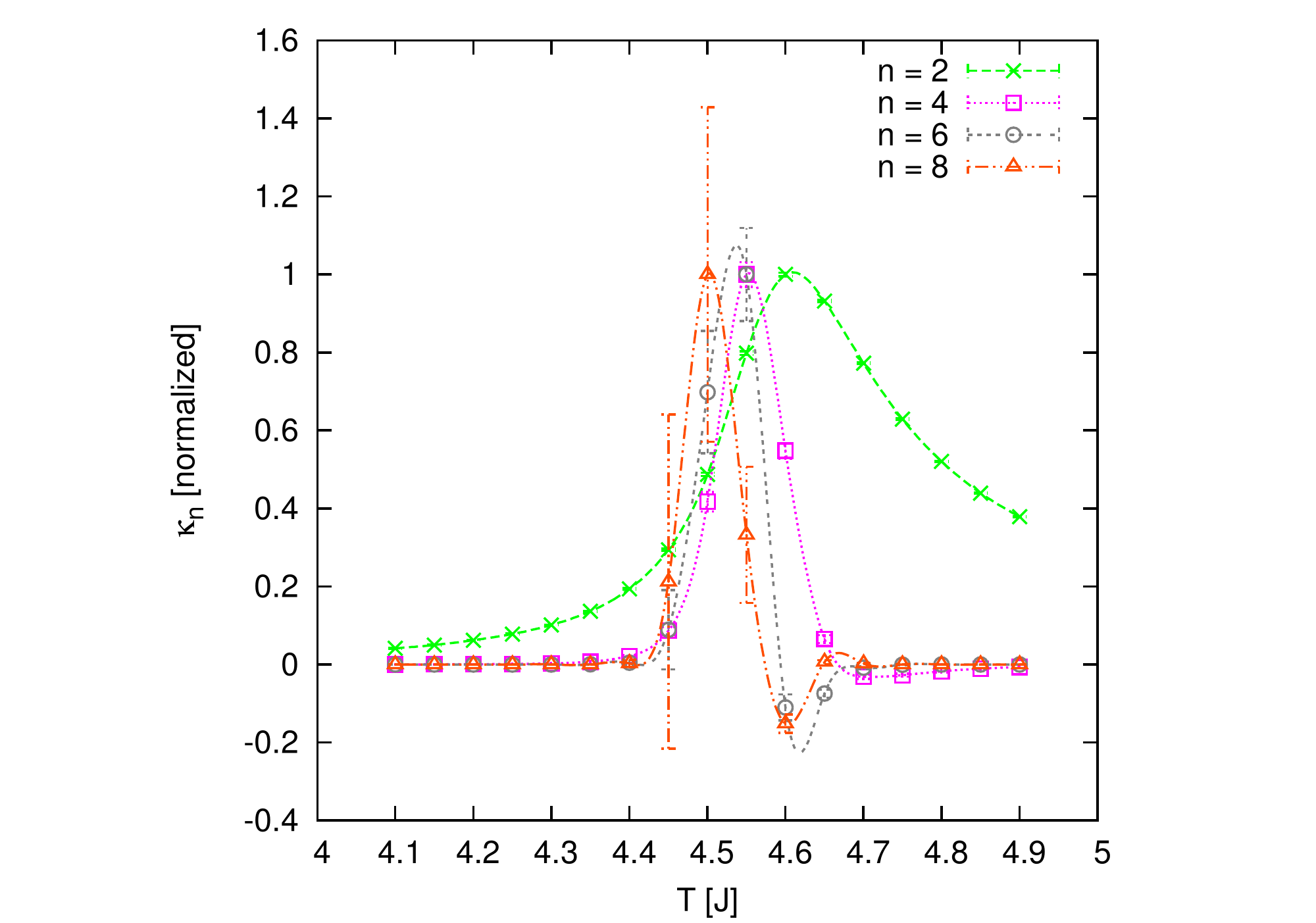}}
\subfigure[$~H_0=0.01J$]
{\includegraphics[scale=0.4,angle=270]{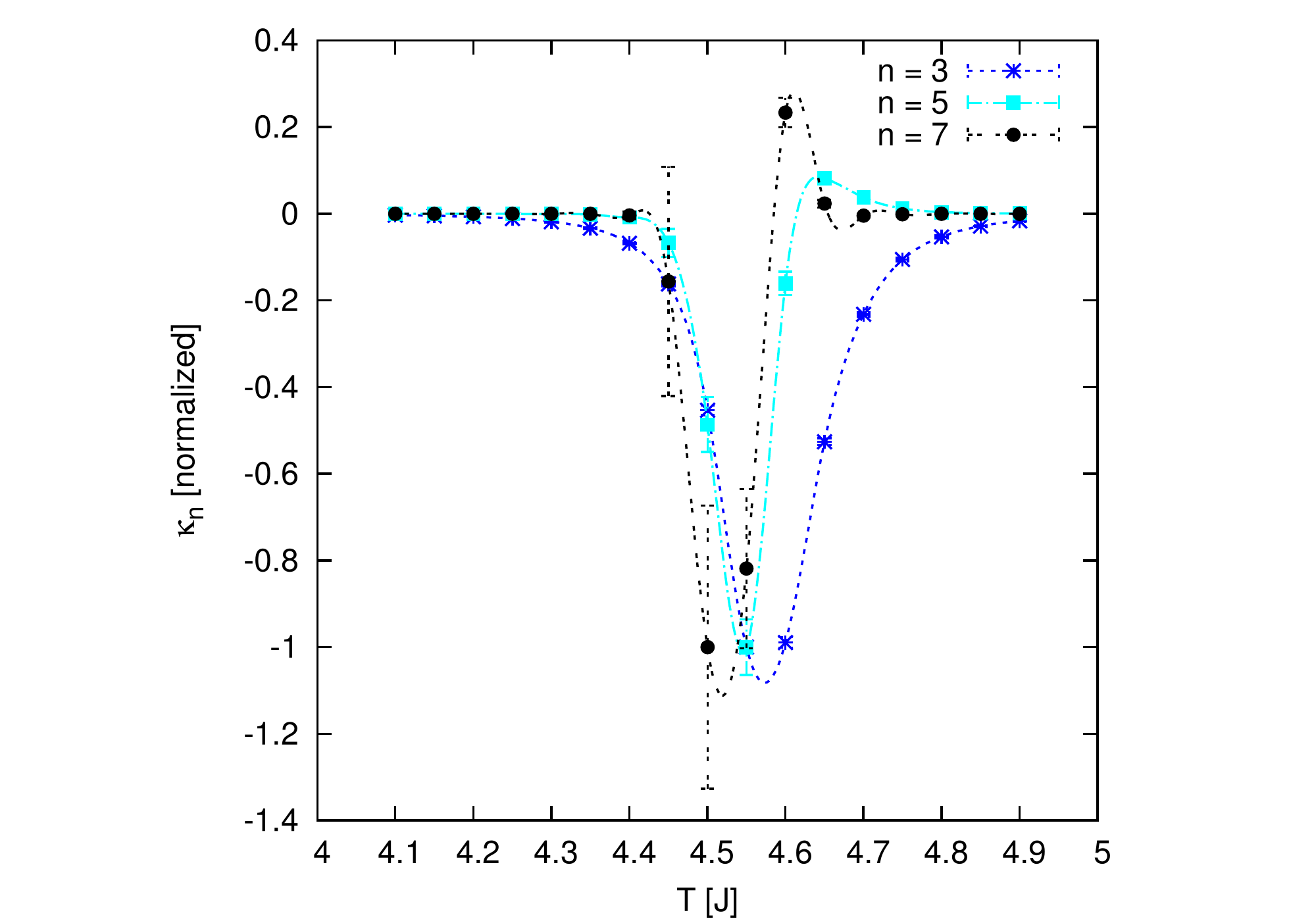}}
\caption{\label{fig:KappaAll}Temperature dependence of the higher order Taylor
coefficients, normalized so that multiple curves can be depicted on each plot.
Even and odd coefficients are separated for clarity.}
\end{figure}

\begin{figure}
\centering
\subfigure[$~H_0=0.1J$]
{\includegraphics[scale=0.4,angle=270]{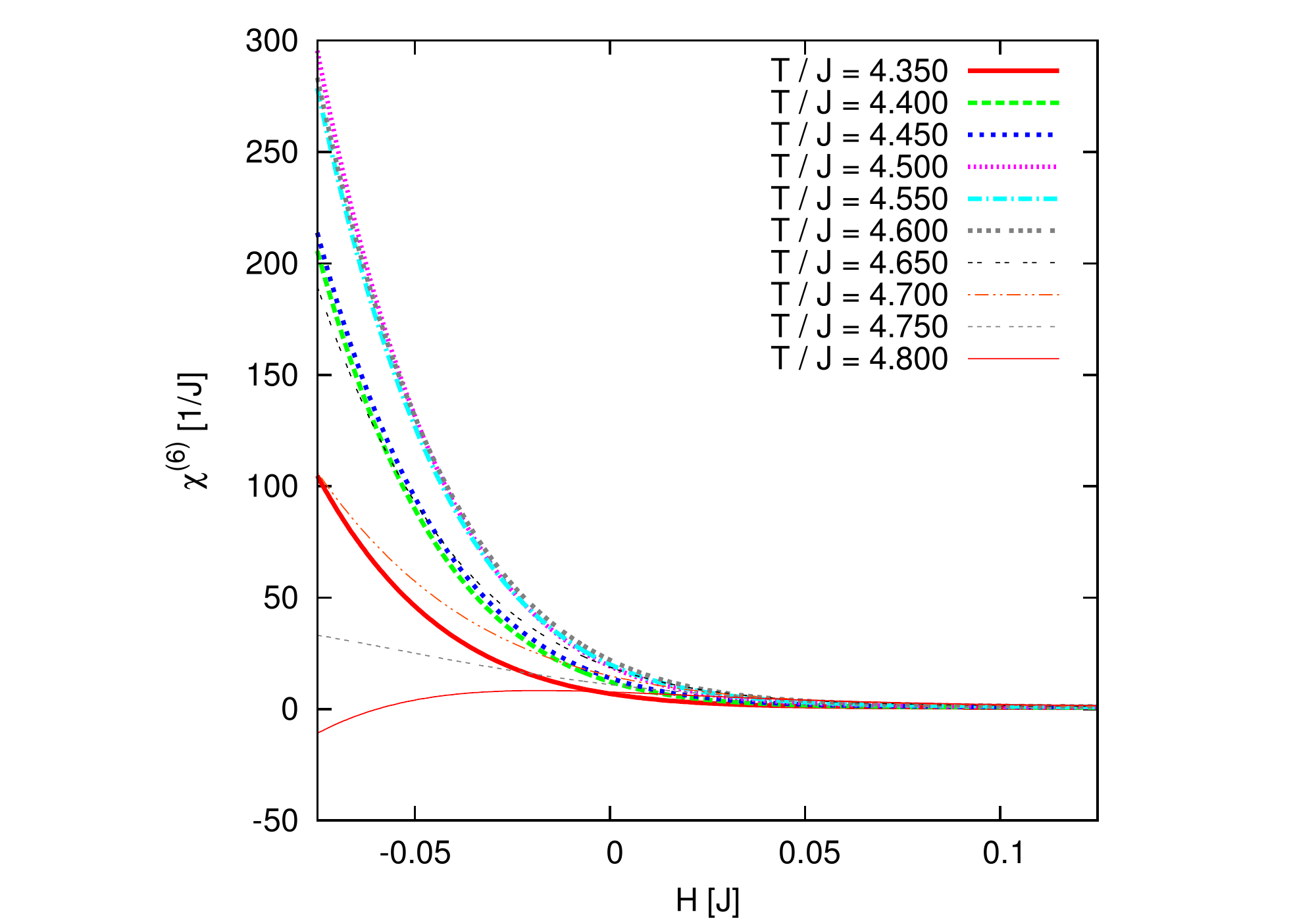}}
\subfigure[$~H_0=0.1J$]
{\includegraphics[scale=0.4,angle=270]{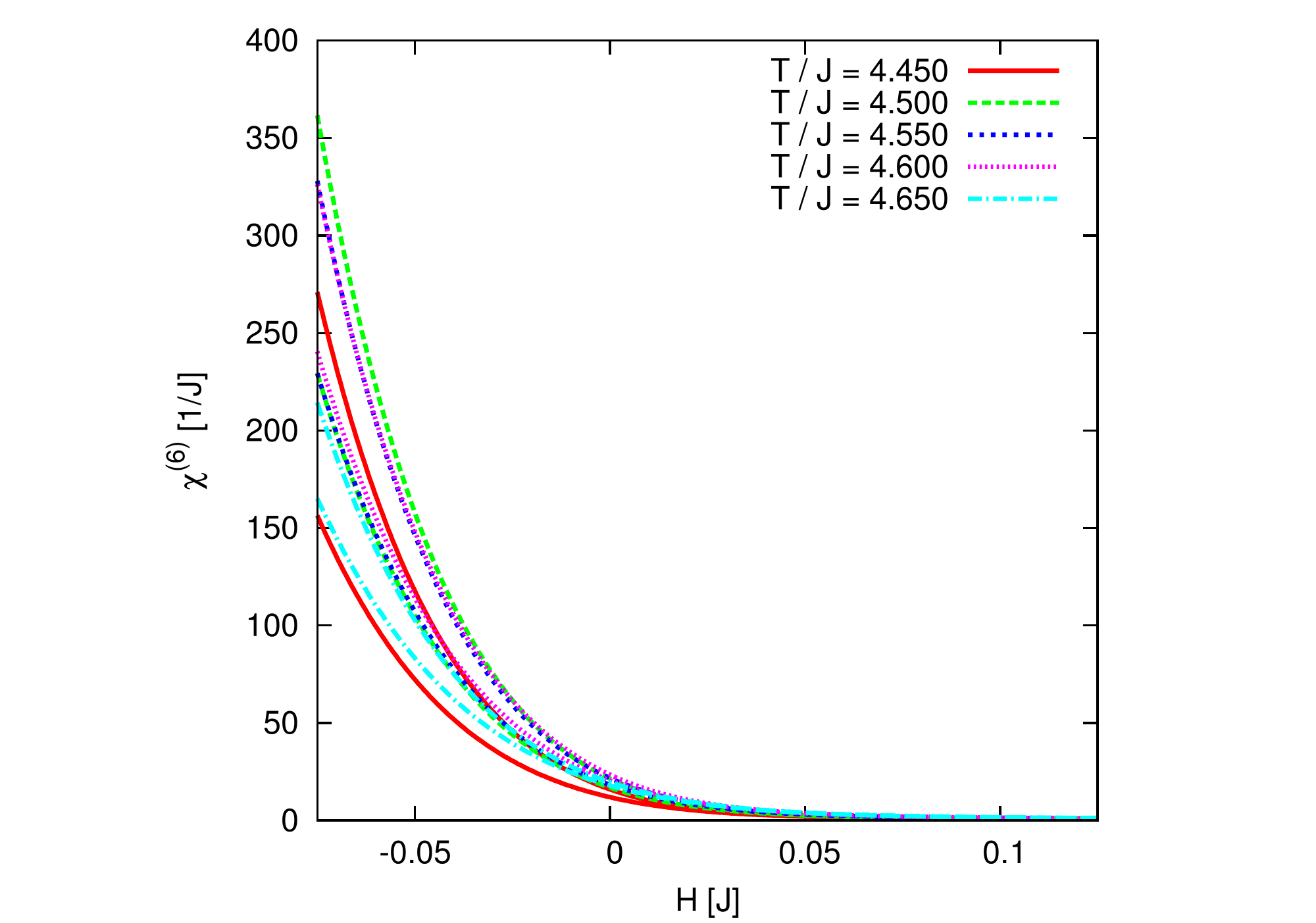}}
\subfigure[$~H_0=0.01J$]
{\includegraphics[scale=0.4,angle=270]{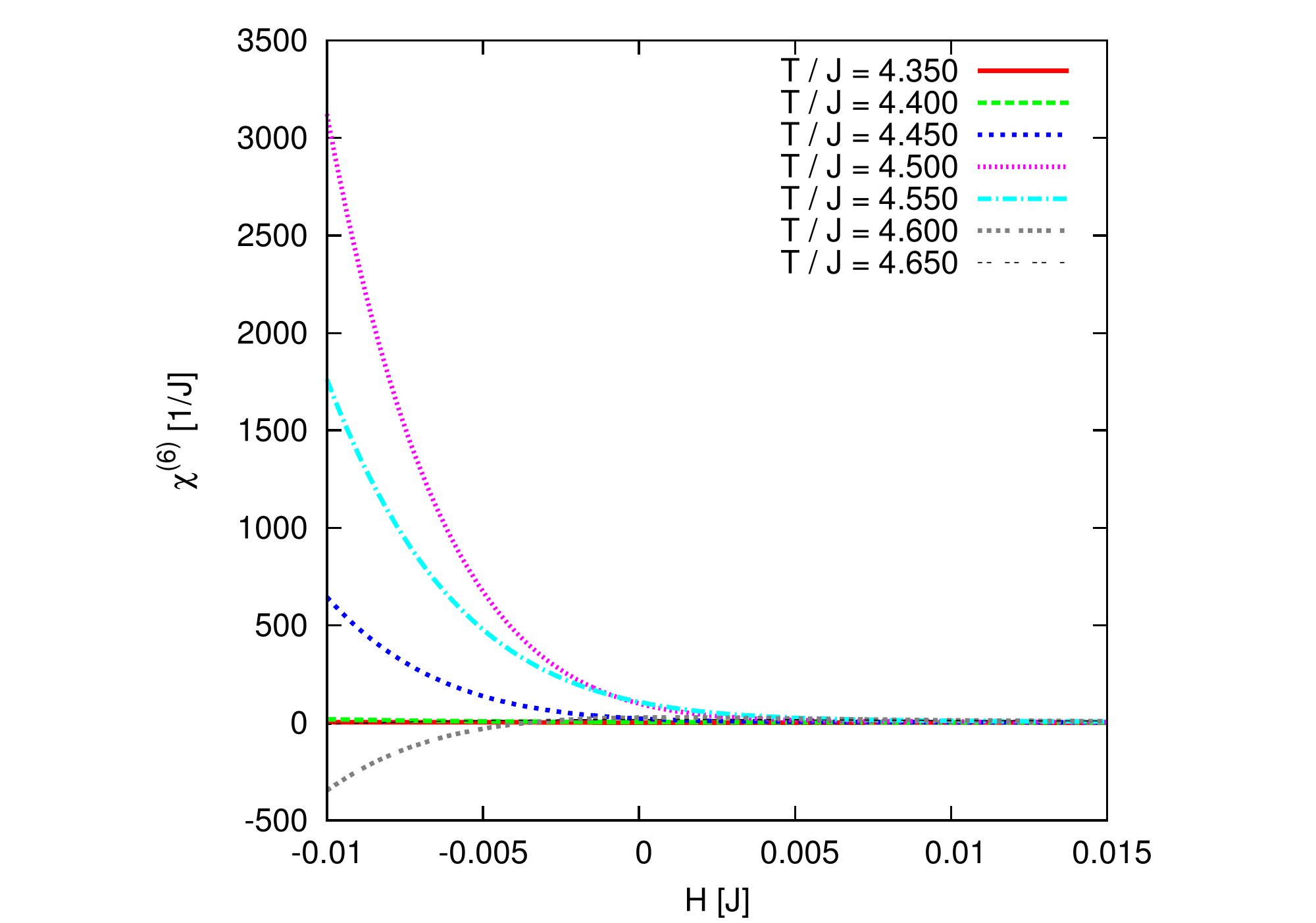}}
\subfigure[$~H_0=0.01J$]
{\includegraphics[scale=0.4,angle=270]{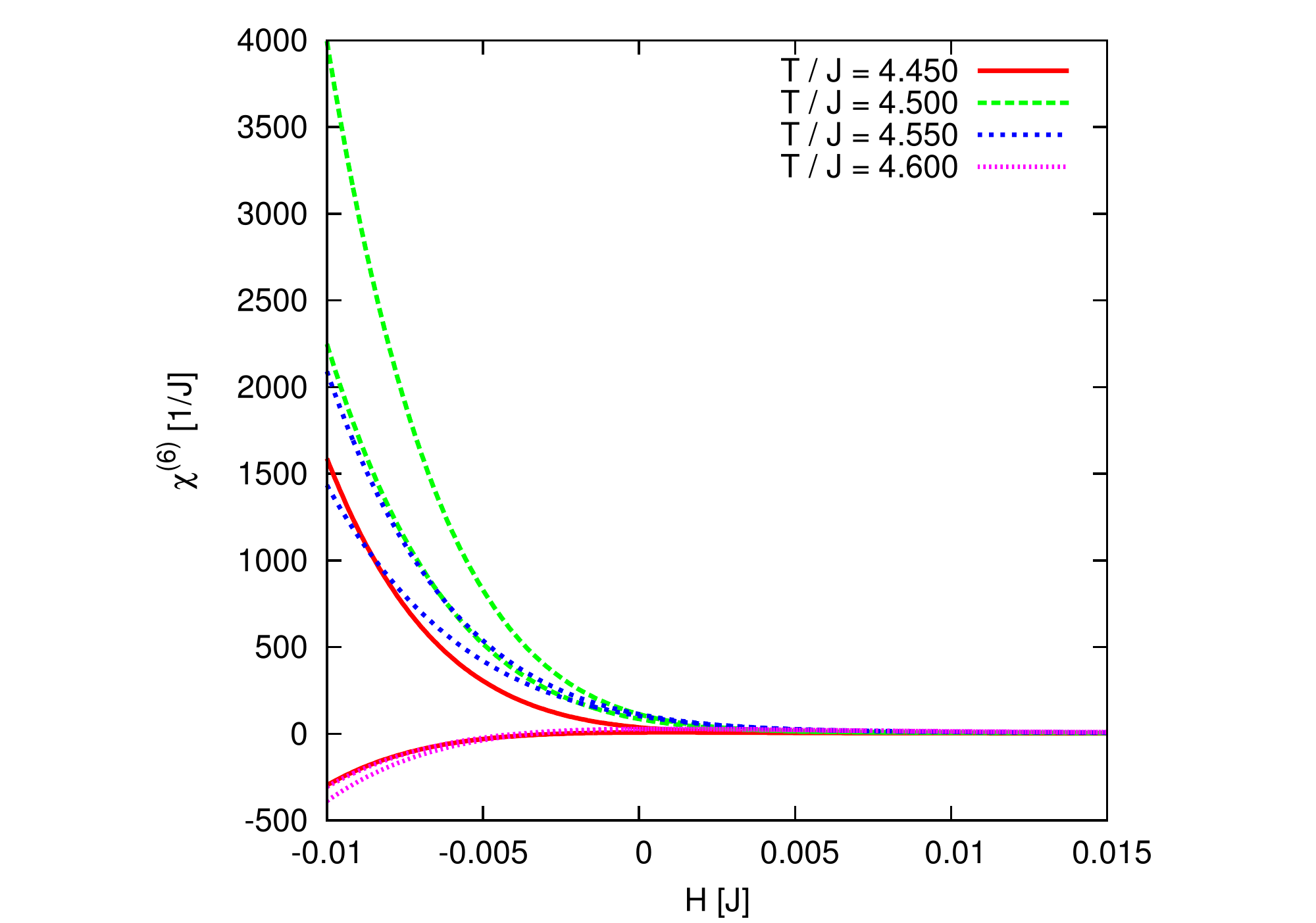}}
\caption{\label{fig:chi6}$\chi^{(6)}$ as defined by Eq. \ref{eq:chi(n)}.
The leftmost graphs depict the progression of $\chi^{(6)}$ throughout the
critical region (uncertainties are not depicted). The rightmost plots
contain the most rapidly divergent graphs indicating that one is very near
$T_c$. Upper and lower curves correspond to uncertainty in $\kappa_8$. When
$H_0=0.1J$ the $T=4.50J$,
$T=4.55J$ and $T=4.60J$ are indistinguishable.}
\end{figure}

\begin{figure}
\centering
\subfigure[$~H_0=0.1J$]
{\includegraphics[scale=0.4,angle=270]{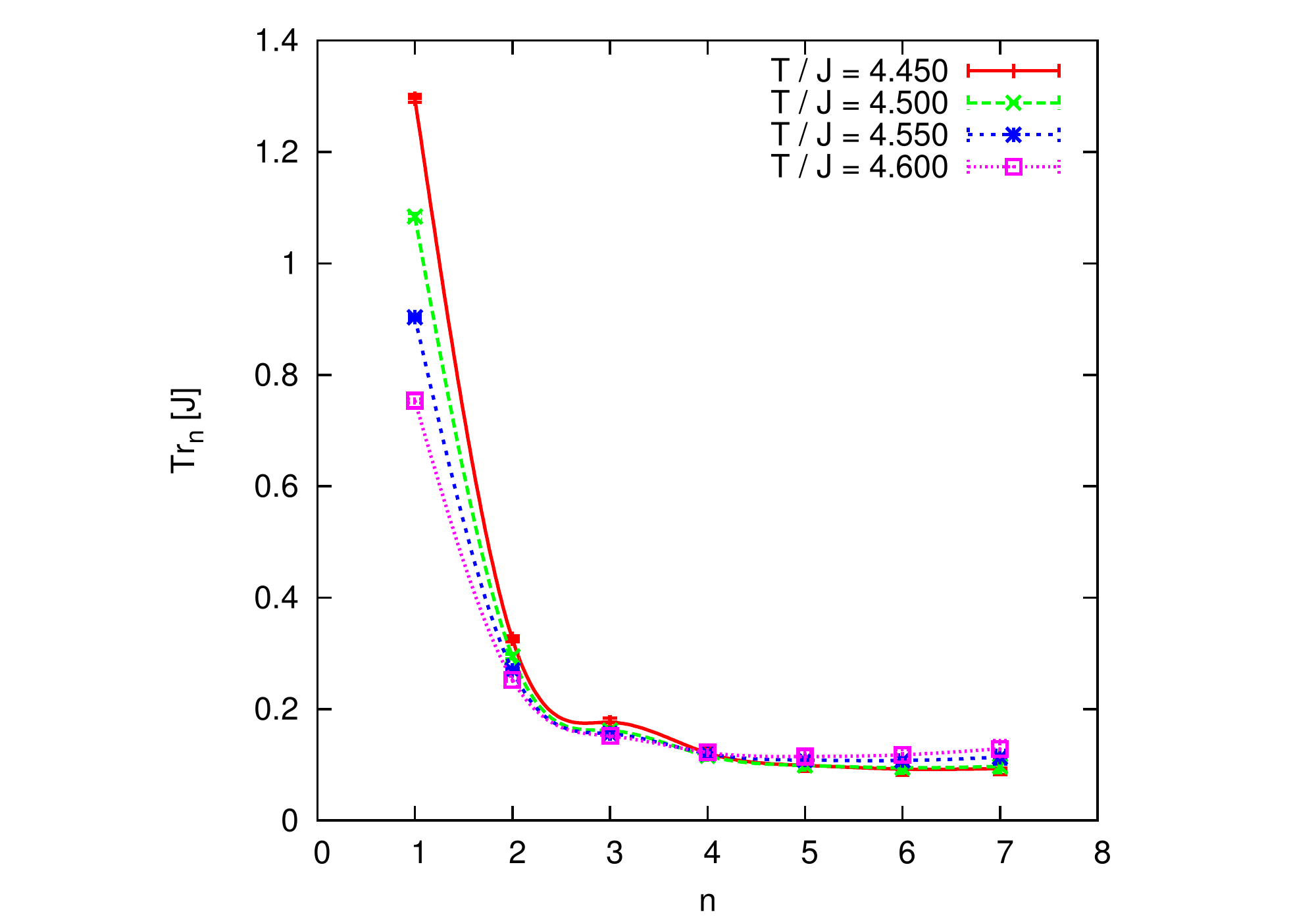}}
\subfigure[$~H_0=0.01J$]
{\includegraphics[scale=0.4,angle=270]{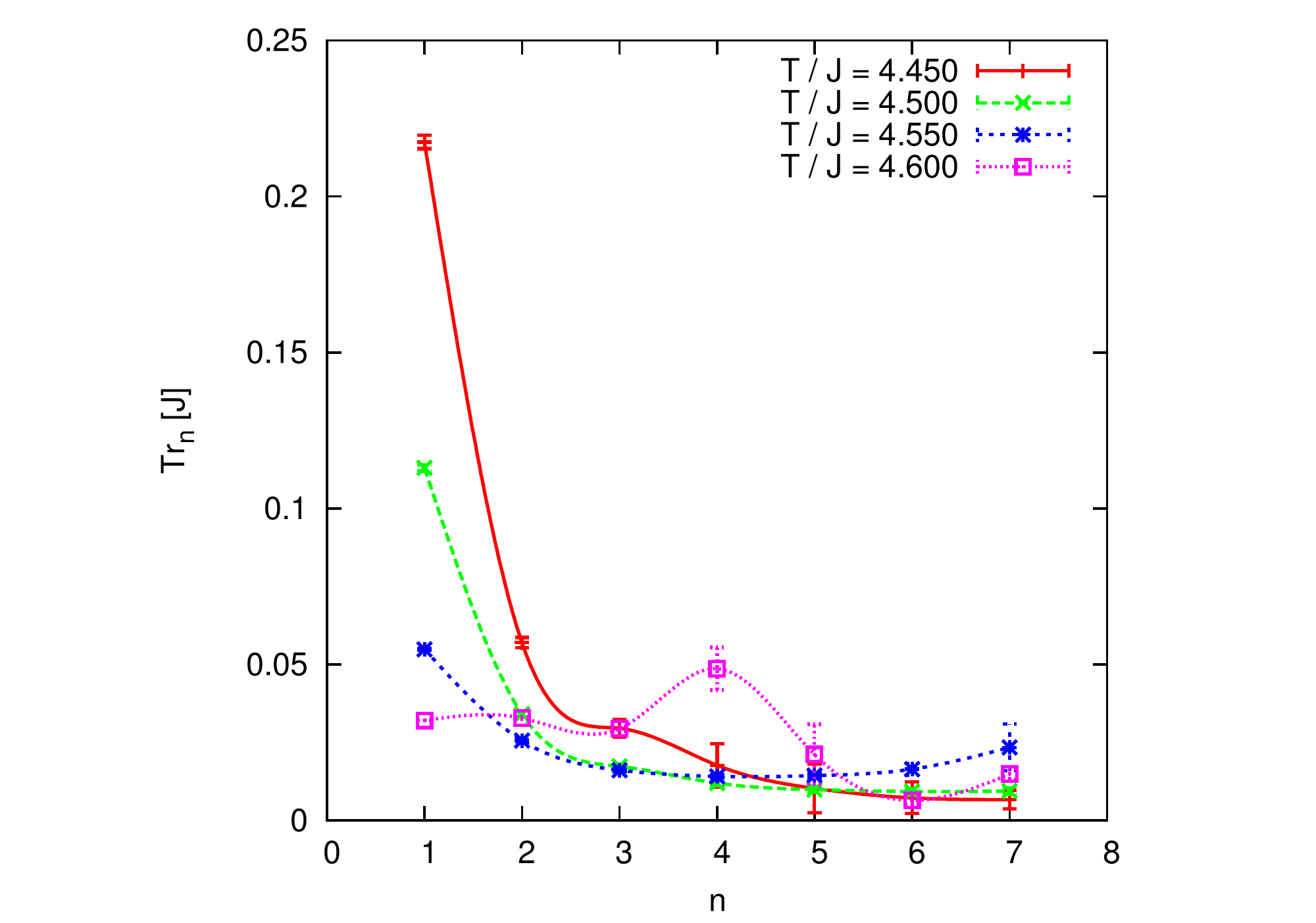}}
\caption{\label{fig:Rn} Convergence of $r_n$ to $r_\infty$ when $T$
is near $T_c$.}
\end{figure}

\begin{figure}
\centering
\subfigure
{\includegraphics[scale=0.36,angle=270]{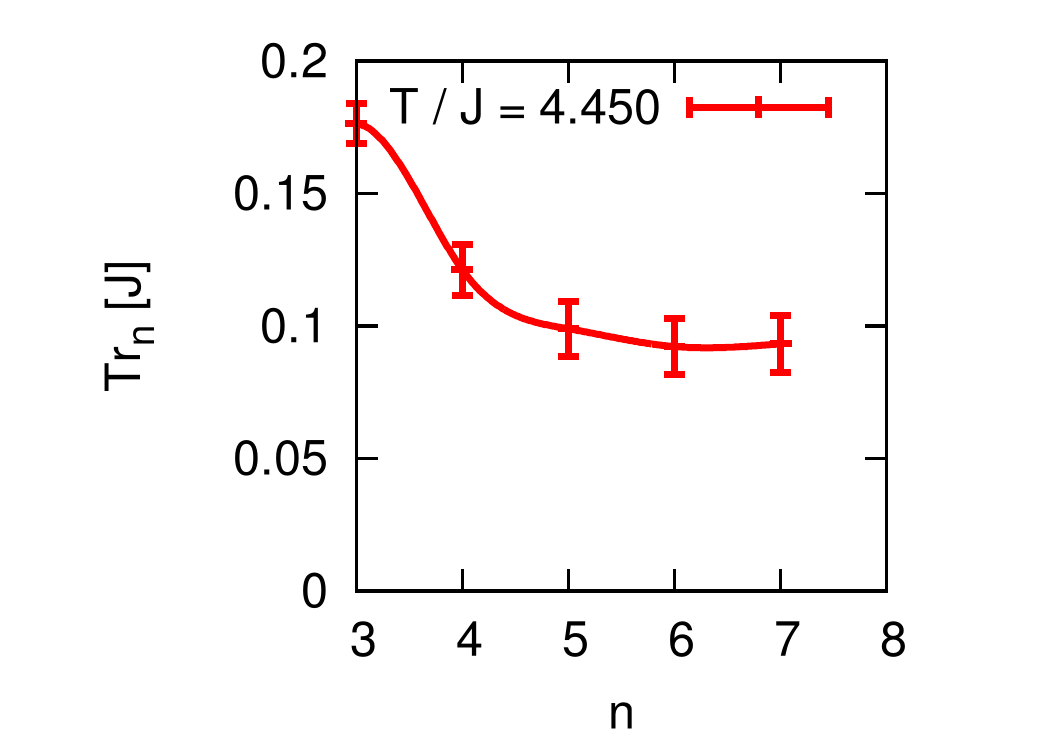}}
\subfigure
{\includegraphics[scale=0.36,angle=270]{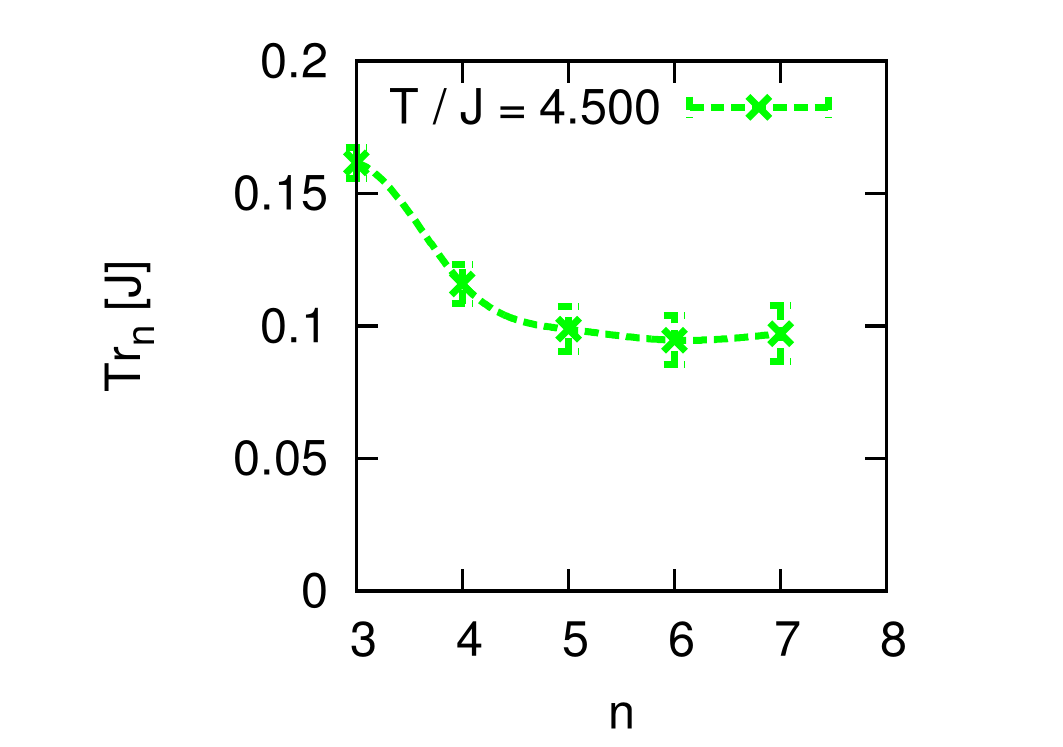}}
\subfigure
{\includegraphics[scale=0.36,angle=270]{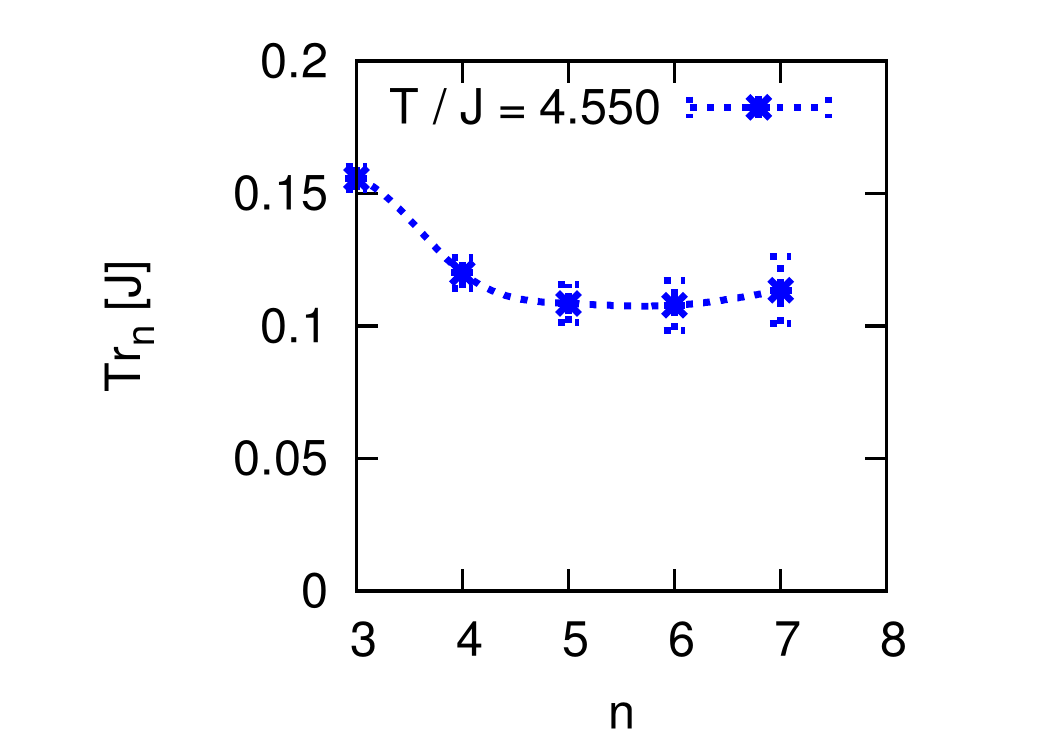}}
\subfigure
{\includegraphics[scale=0.36,angle=270]{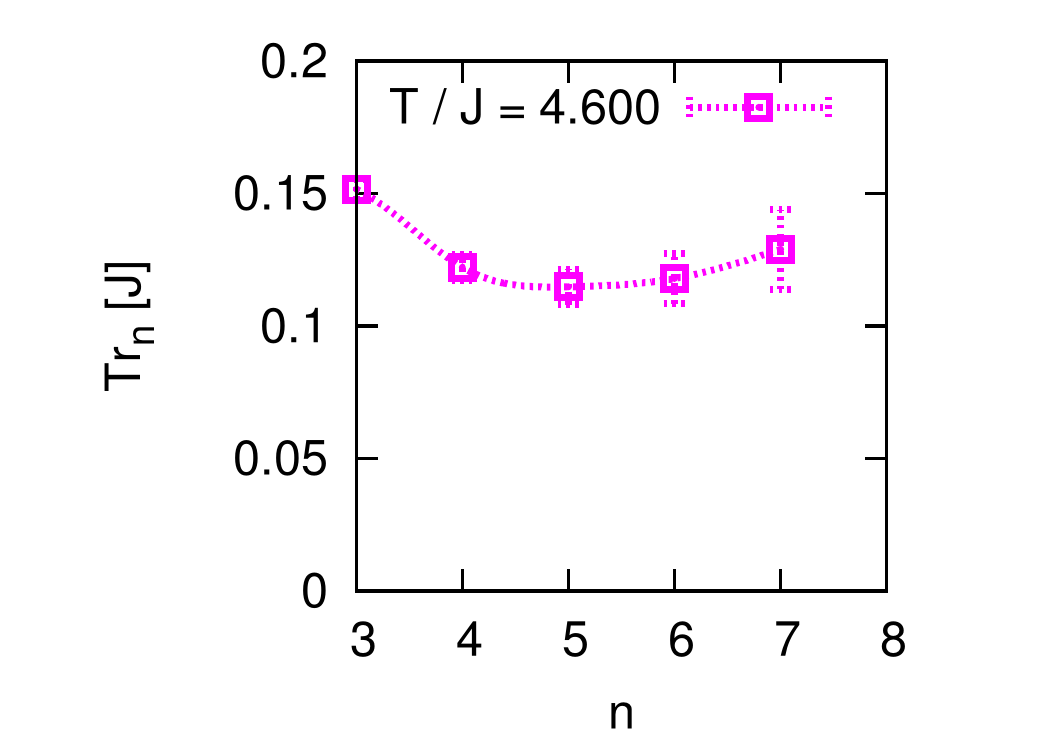}}
\caption{\label{fig:H01Rn}Close-up of the $H_0=0.1J$ graphs in Fig.
\ref{fig:Rn}.}
\end{figure}

\begin{figure}
\centering
\subfigure
{\includegraphics[scale=0.36,angle=270]{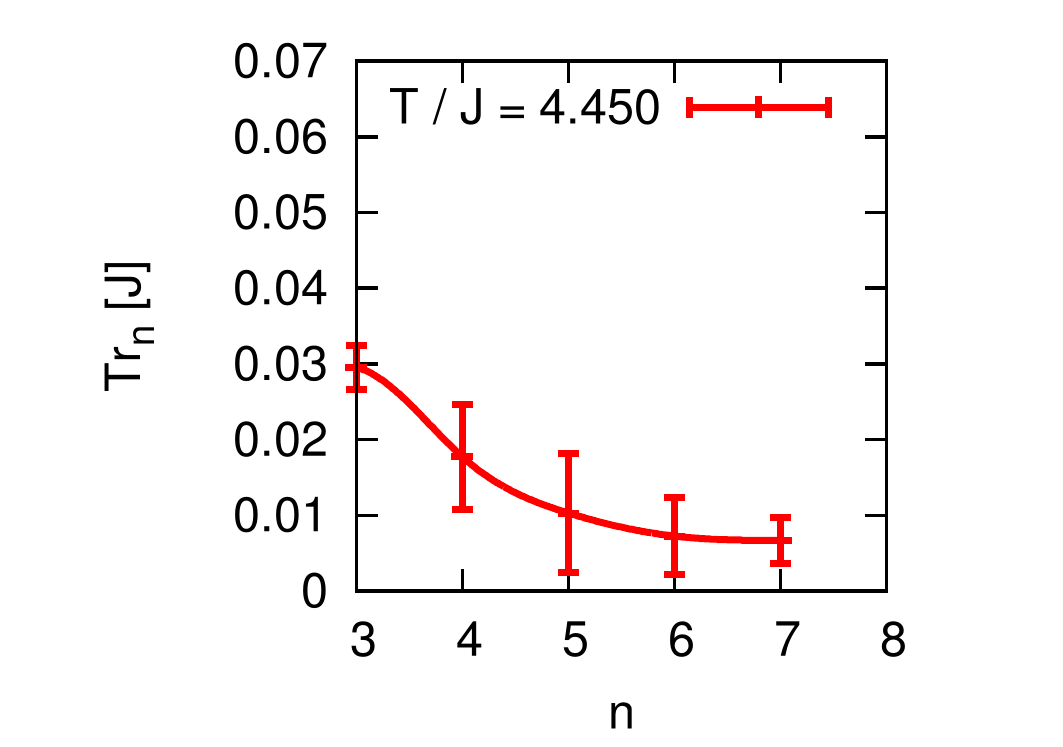}}
\subfigure
{\includegraphics[scale=0.36,angle=270]{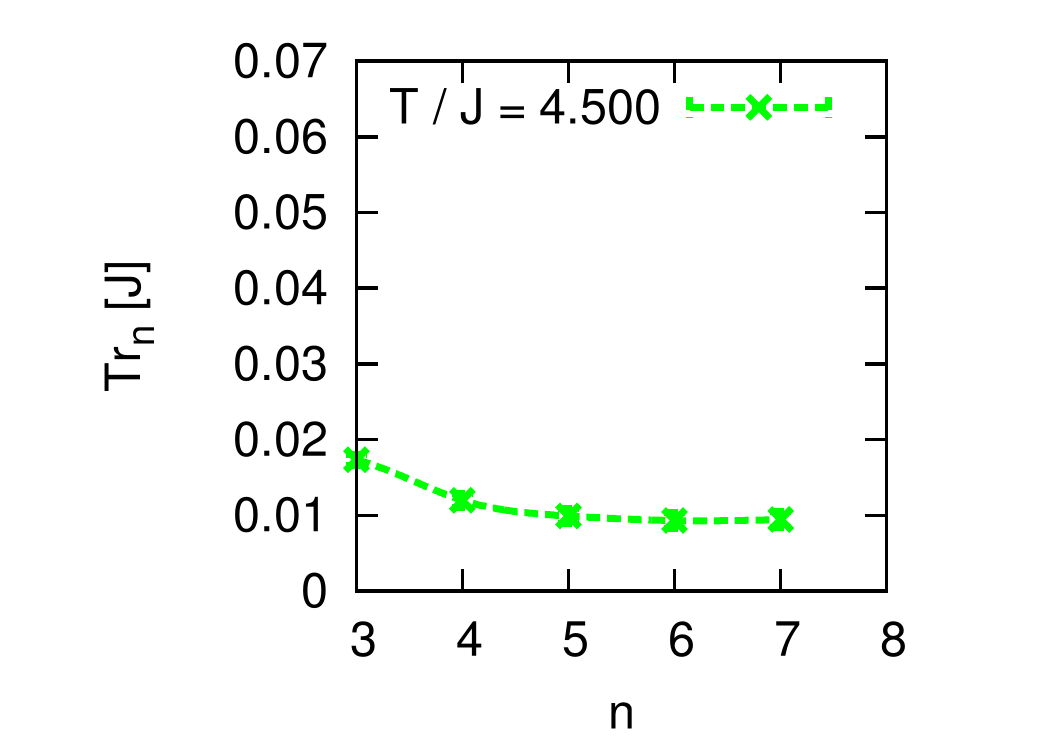}}
\subfigure
{\includegraphics[scale=0.36,angle=270]{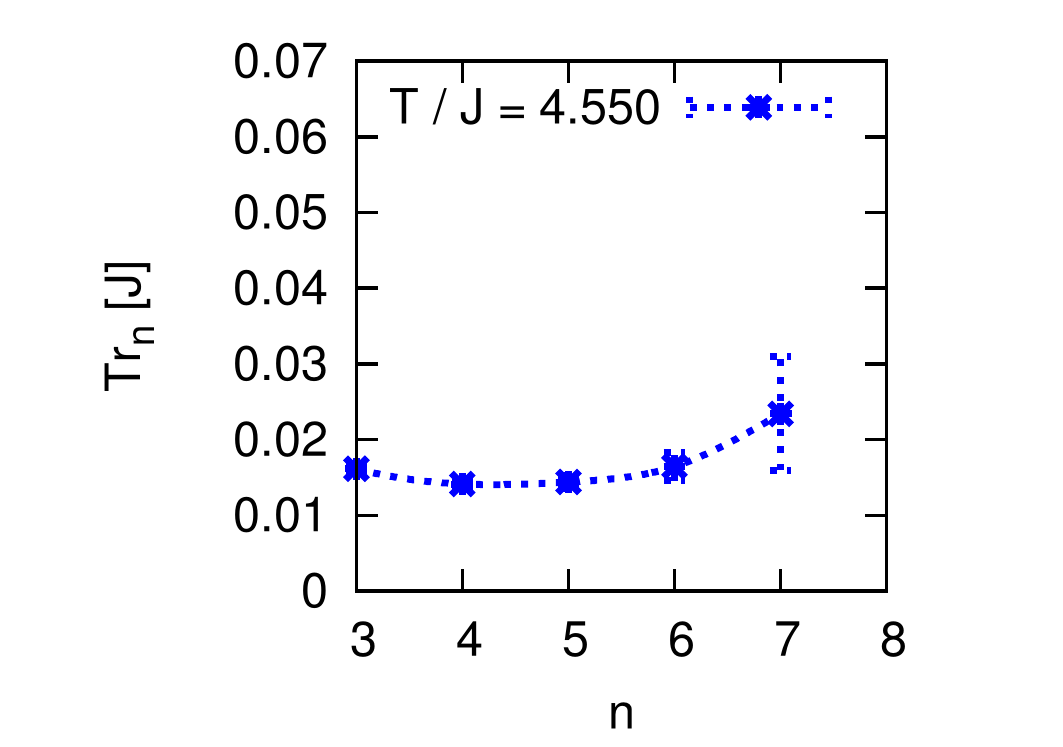}}
\subfigure
{\includegraphics[scale=0.36,angle=270]{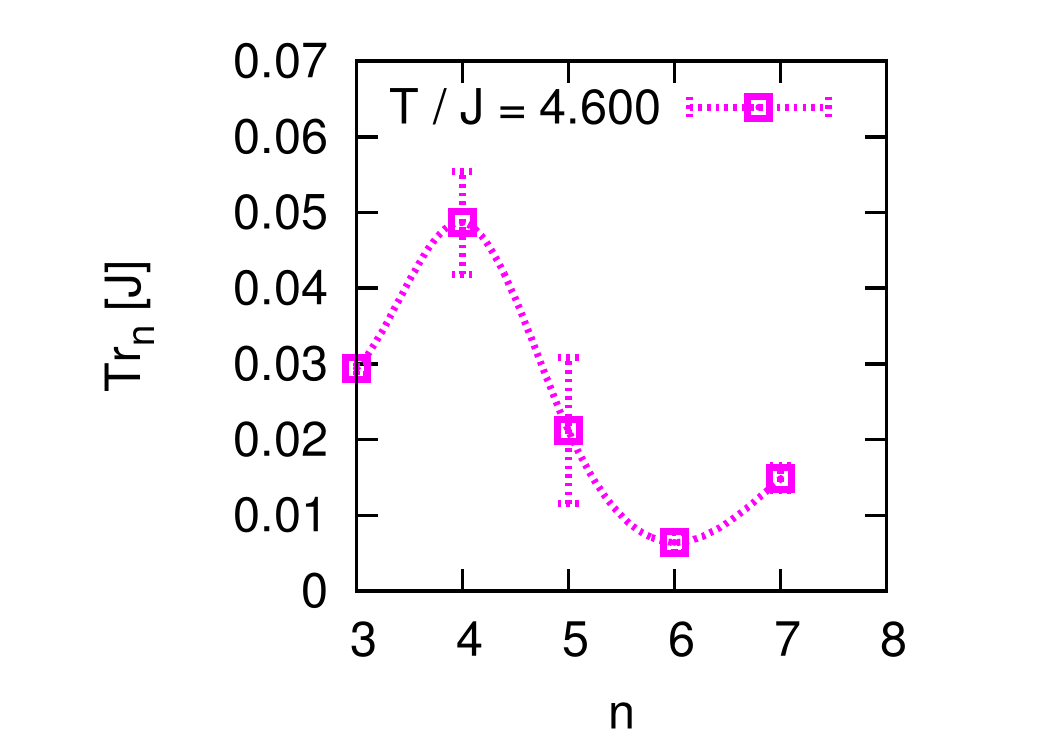}}
\caption{\label{fig:H001Rn} Close-up of the $H_0=0.01J$ graphs in Fig.
\ref{fig:Rn}.}
\end{figure}

\begin{figure}
\centering
\subfigure[$~H_0=0.1J$]
{\includegraphics[scale=0.4,angle=270]{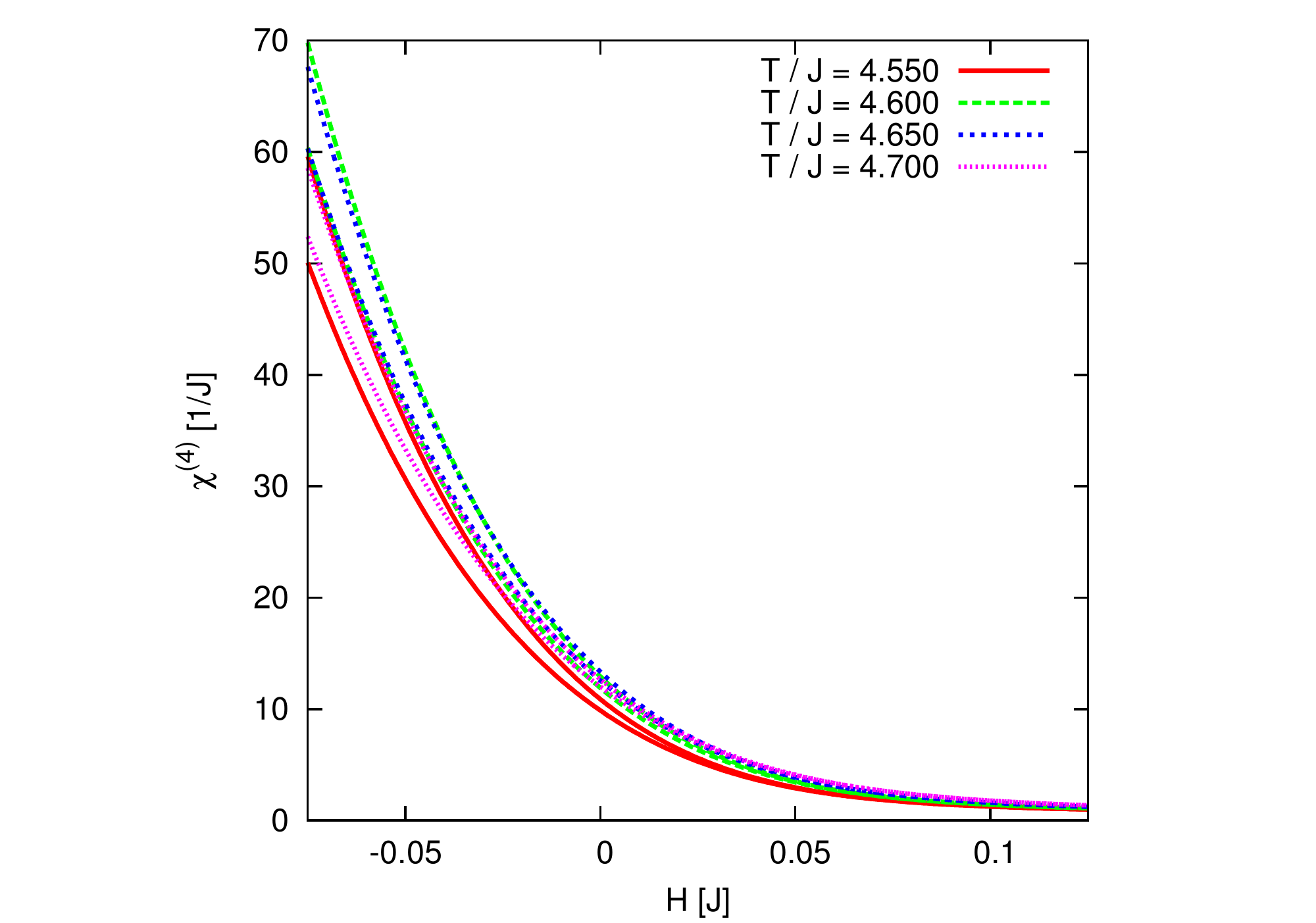}}
\subfigure[$~H_0=0.1J$]
{\includegraphics[scale=0.4,angle=270]{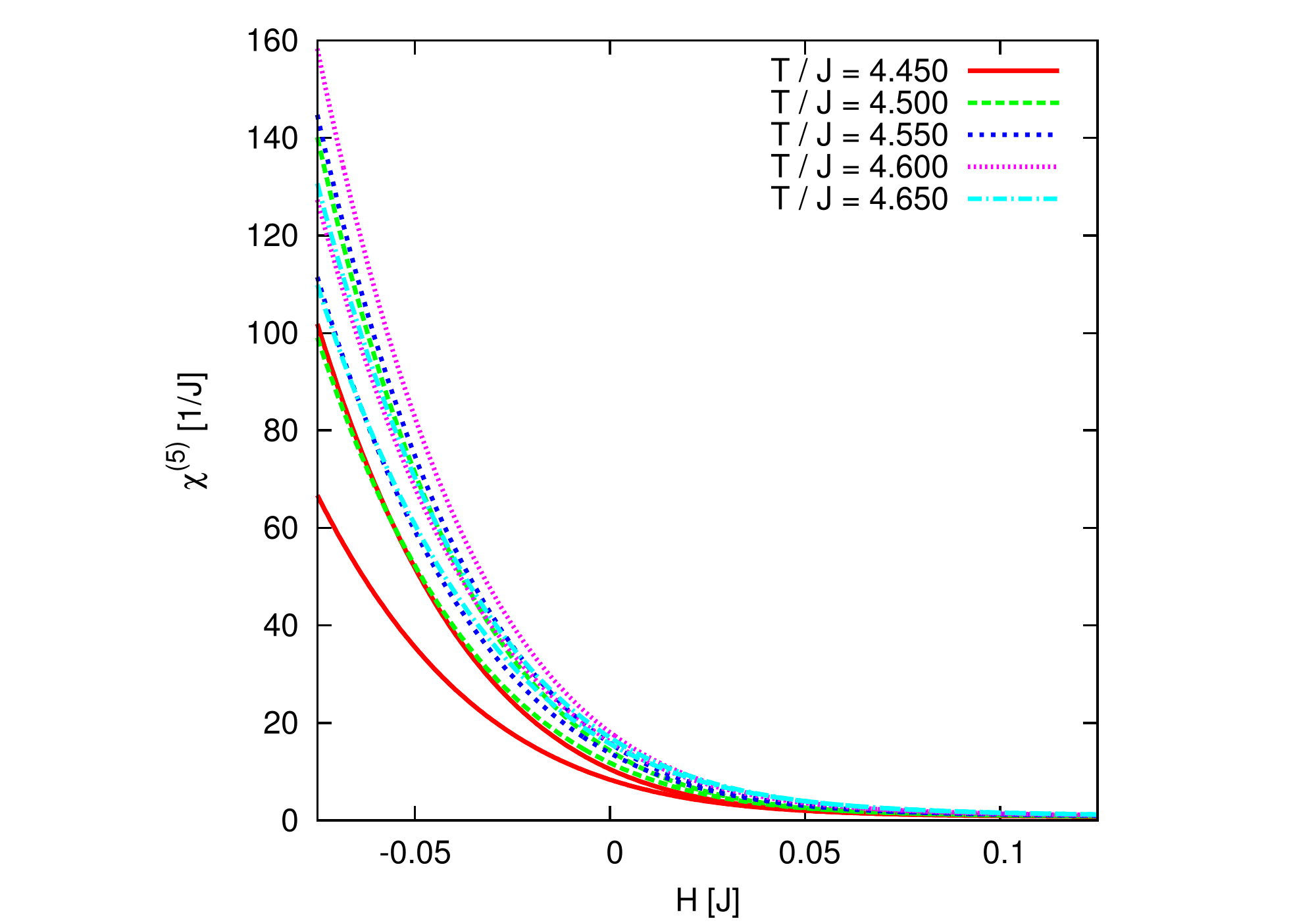}}
\caption{\label{fig:chi45}$\chi^{(4)}$ and $\chi^{(5)}$. When restricted to
lower orders, we see that one would be inclined to overestimate $T_c$.}
\end{figure}

\end{document}